\documentclass[journal]{IEEEtran}
\usepackage{algorithm,algorithmic,amsbsy,amsmath,amssymb,epsfig,bbm,mathrsfs,fancyhdr,fancyvrb,subfigure,url,cite,multirow,xcolor}
\usepackage{graphicx}
\usepackage{epstopdf}
\usepackage{setspace}
\usepackage{array}
\allowdisplaybreaks

\newcommand{\beq}{\begin{equation}}
\newcommand{\eeq}{\end{equation}}
\newcommand{\beqn}{\begin{eqnarray}}
\newcommand{\eeqn}{\end{eqnarray}}

\usepackage{amsmath}

\providecommand{\theoremname}{\textbf{Theorem}}
\providecommand{\propositionname}{\textbf{Proposition}}
\providecommand{\remarkname}{\textbf{Remark}}
\providecommand{\lemmaname}{\textbf{Lemma}}
\providecommand{\corollaryname}{\textbf{Corollary}}
\providecommand{\Definition}{\textbf{Definition}}

\begin{document}

\title{Applications of Auction and Mechanism Design in Edge Computing: A Survey}
\author{Houming Qiu, Kun Zhu,~\IEEEmembership{Member,~IEEE}, Nguyen Cong Luong, Changyan Yi,~\IEEEmembership{Member,~IEEE}, \\Dusit Niyato,~\IEEEmembership{Fellow,~IEEE} and Dong In Kim~\IEEEmembership{Fellow,~IEEE}
\thanks{H. Qiu, K. Zhu, C. Y. Yi are with the College of Computer Science and Technology, Nanjing University of Aeronautics and Astronautics, Nanjing 210016, China (email: \{hmqiu56, zhukun, changyan.yi\}@nuaa.edu.cn).}
\thanks{N. C. Luong is with the Faculty of Computer Science, PHENIKAA University, Hanoi 12116, Vietnam (email: luong.nguyencong@phenikaa-uni.edu.vn).}
\thanks{D. Niyato is with School of Computer Science and Engineering, Nanyang Technological University, Singapore 639798 (email: dniyato@ntu.edu.sg).}
\thanks{D. I. Kim is with the Department of Electrical and Computer Engineering, Sungkyunkwan University, Suwon 16419, South Korea (email: dikim@skku.ac.kr).}
}\maketitle

\begin{abstract}
Edge computing as a promising technology provides lower latency, more efficient transmission, and faster
speed of data processing since the edge servers are closer to the user devices. Each edge server with
limited resources can offload latency-sensitive and computation-intensive tasks from nearby user devices.
However, edge computing faces challenges such as resource allocation, energy consumption, security and
privacy issues, etc. Auction mechanisms can well characterize bidirectional interactions between edge
servers and user devices under the above constraints in edge computing. As demonstrated by the existing
works, auction and mechanism design approaches are outstanding on achieving optimal allocation strategy
while guaranteeing mutual satisfaction among edge servers and user devices, especially for scenarios with
scarce resources. In this paper, we introduce a comprehensive survey of recent researches that apply
auction approaches in edge computing. Firstly, a brief overview of edge computing including three common
edge computing paradigms, i.e., cloudlet, fog computing and mobile edge computing, is presented. Then,
we introduce fundamentals  and backgrounds of auction schemes commonly used in edge computing systems.
After then, a comprehensive survey of applications of auction-based approaches applied for edge
computing is provided, which is categorized by different auction approaches. Finally, several open challenges and promising research directions are discussed.
\end{abstract}


\section{Introduction}
\IEEEPARstart{W}{ith} the rapid development of big data, Internet of Things (IoT)~\cite{A.Galanopoulos2020},
artificial intelligence (AI)~\cite{G.Dandachi2020}, 5G and other intellectual technologies, massive amounts
of data and service requests will be generated at the end devices~\cite{X.Wang2020}. According to the recent
report from Gartner, more than half of the enterprise data will be generated in the edge of the network
rather than the traditional data center (e.g., cloud platform) by 2022. Cloud computing (CC) as
a centralized computing paradigm offers services for end users by migrating data, computation, and storage to
the remote cloud data center. However, massive long-distance data transmission will inevitably cause delay
and network congestion. It indicates that CC cannot meet the increasing requirements for low latency and high
quality of experience (QoE) application scenarios, especially in Internet of Vehicles
(IoV)~\cite{Z.Ning2019}, intelligent networks~\cite{Y.Zuo2020}, telemedicine~\cite{C.Dilibal2020}, smart
city~\cite{D.Zhang2019}, AR/VR~\cite{B.Krogfoss2020}, etc.

The emerging edge computing (EC) provides an effective solution to overcome the limitations of the cloud.
Different from CC, EC is a type of decentralized computing paradigm, which moves data, computation, and
storage from the data center to edge nodes of the network~\cite{W.Shi2016}. Therefore, EC can bring
faster speed of data processing, more efficient transmission, and lower latency since the edge nodes are
closer to the user devices. In addition, EC also provides more intelligent analysis and processing services
near the data sources, e.g., user equipments (UEs), intelligent vehicles, etc. In this case, the
communication delay can be significantly reduced, the system efficiency can be effectively increased, and
security and privacy of data can also be significantly reinforced~\cite{F.Wang2020}. In recent reports, the
global market size of EC has reached at \$3.5 billion in 2019, and exhibiting a compound annual growth
rate exceeding 37\% from 2020 to 2027~\cite{report01}.

Despite of possessing several advantages, EC raises big issues of resource management. In particular, edge
nodes in EC typically have limited resources, i.e., computing, storage, and network resources, while end
users, i.e., service requesters (SRs), have a rapid growth of computing demands. Thus, one issue is how to
efficiently allocate the resources to the SRs. In addition, EC introduces more service providers (SPs) in the
computing market. Since both SPs and SRs are naturally selfish~\cite{S.Pan2019}, how to motivate both the SPs
and SRs to participate in the market is another issue.

To address such issues, auction theory~\cite{V.Krishna2002} as a popular economic approach has been widely
applied, e.g., in wireless networks~\cite{Y.Zhang2013}. Specifically, auction-based mechanisms are promising
since they can fairly and efficiently allocate limited resources of sellers to buyers in a trading form at
competitive prices. An ideal auction-based mechanism should ensure several desirable properties, e.g.,
truthfulness (TF), budget balance (BB), individual rationality (IR), and economic efficiency
(EE)~\cite{W.Vickrey1961}. In particular, with the EE, the auction-based mechanisms guarantee that the
resources are allocated to the buyers that value them the most. Given those advantages of auction theory,
several works~\cite{Y.Luo2021,C.Kai2020,A.Kiani2017,W.Sun2018,S.Guo2020} have recently adopted auction theory
to solve the resource management in EC. It is inspired from the existing works that different types of
auction methods are suitable for different types of problems in specific application scenarios.

Although there are surveys related to EC such as~\cite{P.Porambage2018,R.Yang2019,W.Yu2018,F.Wang2020}, they
do not focus on auction approaches. Also, there is one survey related to auction theory,
i.e.,~\cite{U.Habiba2018}, but it does not focus on EC. To the best of our knowledge, there is no survey
specifically discussing the use of auction theory, an emerging approach, for EC. This motivates us to deliver
the survey with the comprehensive literature review on the auction and mechanism design approaches in EC.


The rest of this paper has the following organization. In Section \uppercase\expandafter{\romannumeral2}, we
present a brief overview of EC and the comparison of cloudlet, fog computing (FC), and mobile edge computing (MEC). Section \uppercase\expandafter{\romannumeral3} gives a review of auction theory. Section \uppercase\expandafter{\romannumeral4} presents the applications of auction approaches for EC. Section \uppercase\expandafter{\romannumeral5} highlights open challenges and future research
directions of utilizing auction approaches to EC. Finally, Section \uppercase\expandafter{\romannumeral6}
concludes for this paper.

Some important definitions of the acronyms that will be frequently used are summarized
in Table~\uppercase\expandafter{\romannumeral1}.

\begin{table}
\caption{Summary of common used in this paper}
\label{table1}
\small
\begin{tabular}{p{55pt}p{170pt}}
\hline\noalign{\smallskip}
Acronym & Definition \\
\noalign{\smallskip}\hline\noalign{\smallskip}
BB   & Budget balance \\
CE   & Computational efficiency\\
EE   & Economic efficiency\\
EC   & Edge computing \\
ECS &  Edge computing server \\
FC   & Fog computing \\
FCN  & Fog computing node \\
IR   & Individual rationality\\
IC   & Incentive compatibility\\
MEC  & Mobile edge computing\\
MDs  & Mobile devices \\
MUs  & Mobile users \\
QoS  & Quality of server\\
QoE  & Quality of experience\\
SRs  & Service requesters \\
SPs  & Service providers \\
TF   & Truthfulness\\
UEs  & User equipments\\
\hline
\end{tabular}
\end{table}

\section{Fundamentals of Edge Computing}
In this section, we first give an overview of EC, including three main types of computing
architectures, i.e., cloudlets, FC, and MEC. Then, we present the comparison of cloudlet, FC, and MEC, and
discuss the advantages and disadvantages of each computing architecture.

\subsection{Main Paradigms of Edge Computing}
\begin{figure}[!t]
\begin{center}
\includegraphics[width=3.3in]{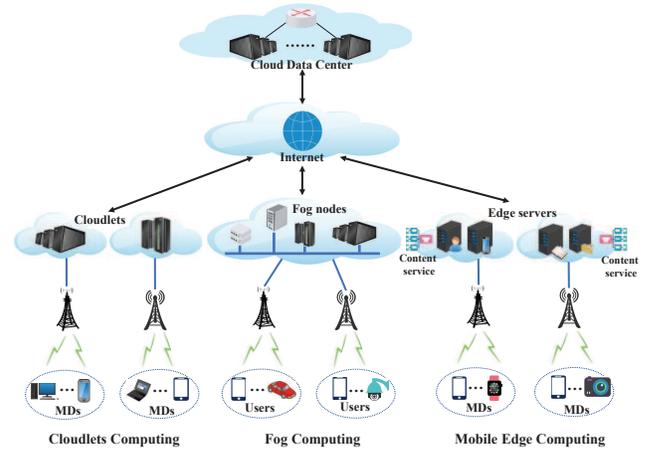}
\caption{The system architecture of EC, include three main paradigms, i.e., cloudlet, FC and MEC.}
\label{fig:ThreeTypeOfEC}
\end{center}
\end{figure}

As an emerging computing framework, EC has been attracting great attentions because it can provide
ultra-low-latency services for end devices. As illustrated in Fig.~\ref{fig:ThreeTypeOfEC}, the system
architecture of EC includes three different paradigms, i.e., cloudlet, FC, and MEC. In the
following subsections, we will give a brief overview of each computing paradigm.

\subsubsection{Cloudlet Computing}
\

\noindent

Cloudlet~\cite{M.Satyanarayanan2009} is a computing architecture combining both mobile computing and CC.
Enlightened by the definition in~\cite{M.Satyanarayanan2009}, cloudlet is a trusted cluster of computers that
provides cloud services at the edge of the network. In general, the cloudlet acts as the middle layer in the
three-tier architecture, i.e., mobile devices (MDs), cloudlets and the cloud. A cloudlet can usually be
regarded as "data center in a box", deployed near mobile devices to support offloading and caching.
Furthermore, the ideal location of the cloudlet is at the edge of the network which can offer an one-hop
access with high bandwidth for MDs. Through cloudlets, MDs can offload latency-sensitive or
computing-intensive applications to achieve shorter latency and less system overhead. In fact, it is obvious
that the cloudlet enhances the capability of mobile cloud computing (MCC)~\cite{H.T.Dinh2013,M.A.Marotta2015}
in addressing the latency challenge between MDs and the cloud.

\subsubsection{Fog Computing}
\

\noindent

As a decentralized computing architecture, FC was introduced by CISCO in 2012 to break
through the limitation of the CC services at the edge of networks~\cite{F.Bonomi2012}. The main
idea of FC is to migrate large-scale processing tasks from the CC center to edge servers
that are close to end devices~\cite{L.M.Vaquero2014}. Compared with the CC architecture, a set
of medium-sized computing units gathering as fog layers are placed between edge devices and the cloud, where
a single computing unit is defined as a fog computing node (FCN). In general, an FCN can provide a set of
medium-sized services, (e.g., computing, storage, networking) to edge devices. In FC, shortening the distance
between data processing unit and data source both physically and logically is the core. As large-scale
processing tasks are mostly carried out in FCNs, the latency of task processing and the network transmission
load are greatly reduced. Due to the locality of FC, end devices can obtain various benefits such as
real-time transmission of data, pre-analysis of data source, etc. However, the decentralized architecture of
FC may cause security and privacy leakage
issues~\cite{C.Mouradian2018,A.Bandyopadhyay2020,P.Kayal2019,Y.Jiao2019,X.Peng2020,Y.Zhang2019}.
Generally, FC is more suitable for IoT systems.

\subsubsection{Mobile Edge Computing}
\

\noindent

In 2014, European Telecommunications Standards Institute (ETSI) first proposed the concept of MEC~\cite{Y.C.Hu2015}. They suggest to deploys sufficient computing capacity,
storage space and service environments to the edge network within the radio access network (RAN). The main
idea of MEC is to distribute highly complex and heavy computation tasks to adjoining edge servers, providing
ultra-low-latency computing services, higher bandwidth for transmission, less consumption for
energy~\cite{Y.Jararweh2016,C.Kai2020}, etc. Thus, the workload of end users can be greatly alleviated by
offloading highly complex and computing-intensive tasks to edge nodes rapidly. Moreover, the battery
lifetime and storage space of user devices (e.g., IoT devices) can be significantly prolonged and expanded,
respectively. Therefore, user devices can run various latency-sensitive or computing-intensive applications
such as unmanned aerial vehicles~\cite{Z.Ullah2020}, smart city~\cite{D.Zhang2019},
AR/VR~\cite{B.Krogfoss2020}, etc. It is noteworthy that context-awareness, as a key feature of MEC, can
promote and improve context-aware services for user devices. Due to the low concentration and small-scale of
data resources of subscribers, the probability of being attacked is much smaller for edge servers in the MEC.
In addition, many MEC servers are equipped with identity authentication, intrusion detection and data
encryption which can effectively address the security-and-privacy
issues~\cite{T.Salman2019,A.Asheralieva2020,Z.Li.Z2019,Y.Jiao2018,W.Sun2020,K.Xiao2020}.
It indicates that the applications with privacy-sensitive and
security-sensitive can be well supported by MEC and greatly benefit from MEC. Given the aforementioned
advantages, it is no doubt that MEC eliminates the drawbacks of FC and MCC. With the vigorous development of
IoT, more applications will be supported by MEC~\cite{N.Abbas2018,R.Han2020}.

\subsection{Comparison of Cloudlet, FC and MEC}
EC paradigm generally contains three different representation forms, i.e.,
cloudlets~\cite{M.Satyanarayanan2009}, FC~\cite{F.Bonomi2012}, and MEC~\cite{R.Roman2018}. It
is obvious that all of them have the identical idea, i.e., migrating computing resources and services
from the central node to the edge nodes of networks. Nevertheless, there are also some differences between them, such as node location, context awareness, access mechanisms, proximity and inter-node communication.

FC has the most comprehensive inter-node communication support, with multiple access mechanisms. Compared
with FC, MEC and cloudlets only have partial inter-node communication support. It is noteworthy that MEC has
the highest context awareness as a result of obtaining detailed information of end users, e.g., location and
network load. FC is worse than MEC in context-awareness owing to the limited view of the network devices,
e.g., routers, switches, etc. Fortunately, the context-awareness of FC can be effectively improved by the
strong capability of inter-node communication. Furthermore, the standalone architecture of cloudlets leads to
the lowest context-awareness because the devices connected to the cloud are independent of each
other~\cite{K.Dolui2017}. Regarding to the computing time, MEC and cloudlets can be able to respond to
assigned tasks timely due to the allocation strategy and the virtualized property of resources, while the
legacy devices used in FC usually have poor processing and storage capacity which may prolong the computing
time.

Although EC is an efficient solution for processing computing intensive and latency-sensitive tasks of edge
devices, the limited resources of edge servers restrict its application and development. As a popular trading
form, auction can efficiently allocate resources while satisfying the heterogeneous requirements of both SPs
and SRs~\cite{T.Bahreini2018,A.Bandyopadhyay2020,S.Yang2020,Y.Yue2019,S.Guo2020,Q.Xu2018}. Thus, it has
attracted a lot of research interests recently.

\section{Overview of Auction Theory}
As a way to enhance the efficiency of social resource allocation, auction theory incentivizes
commodities or services be purchased by the buyers who need it most, and determines the trading price that
maximizes the total profit of the sellers. Due to many advantages of auction theory, it has widely been
applied to both economics and engineering areas. First of all, we introduce the basic terminologies of
auction theory. Then, some popular auction methods are briefly discussed.

\subsection{Basic Terminologies}
\subsubsection{Seller}
A seller refers to the owner of the auction commodities or services, and wants to sell them at certain prices for gaining maximum profit. In the EC market, edge servers usually act as the sellers which own a number of commodities, such as computing and storage resources, network bandwidths, etc.

\subsubsection{Bidder}
A bidder is a buyer who wants to purchase commodities or services from sellers. In the EC market, UEs and MDs
usually act as buyers that want to purchase diverse resources (e.g., computing resource) to
process computing-intensive and latency-sensitive tasks. Both bidders and sellers are auction participants.

\subsubsection{Auctioneer}
An auctioneer typically plays the role of an executor to implement auction algorithm, and determines the
winners and payments according to the auction rules for both buyers and sellers. In the EC market, the
auctioneer can be a service provider (a seller) or a trusted third party.

\subsubsection{Commodity}
In an auction, a commodity refers to a trading object between a seller and a buyer. Sellers sell a
commodity with a value at an optimal price through the competition across the buyers. In the EC market,
the commodities can be computing resources, cache resources, or network resources.

\subsubsection{Price}
In general, the price emerged in the form of competition during an auction process. It may be an asking
price, i.e., the price that a seller agrees to sell the commodity, a bidding price, i.e., the price that a
buyer agrees to pay for the commodity, or a hammer price (transaction price), i.e., a final trading price in
the auction.

\subsection{Auction Approaches Commonly Used For Edge Computing}
Generally, the EC market typically includes multiple SPs, i.e., sellers, and multiple MDs, i.e., buyers,
where buyers need sellers' resources to process computing intensive and latency-sensitive tasks. Thus, double
auction is commonly applied to address resource trading in the EC market due to its many-to-many
structure~\cite{Y.Yue2018,Y.Yue2019,W.Sun2018,R.Zhang2019,H.Hong2020,Y.Zhang2013,Z.Gao2019}. Also, the MDs in
the EC market usually bid for a bundle of resources (e.g., computing, storage, network, energy, etc.), and the combinatorial auction is an effective solution. Therefore, the following introduces the combinatorial auction and double auction.

\subsubsection{Combinatorial Auction}
A combinatorial auction~\cite{S.deVries2003} is an auction, in which each bidder bids for a combination of
various commodities. Compared with the traditional auction methods, buyers can obtain a package of
combinatorial commodities which contain different types of commodities.

\begin{figure}[!t]
\centering
\includegraphics[width=3.3in]{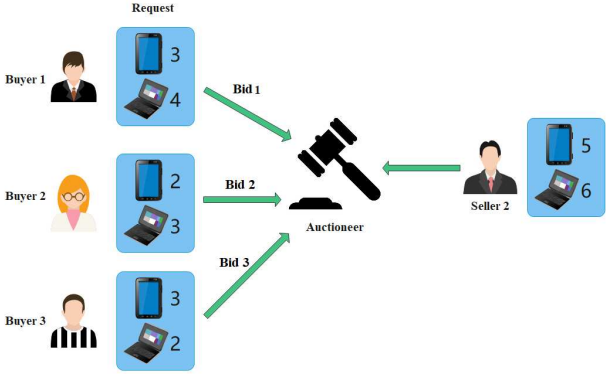}
\caption{The combinatorial auction model.}
\label{fig:combinatorial-auction}
\end{figure}

In the combinatorial auction, each buyer submits a bid to the auctioneer that indicates the demand of a
bundle of commodities rather than a single commodity. After collecting bids/asks submitted by buyers/sellers,
the auctioneer gives an optimal allocation scheme over buyers. Fig.~\ref{fig:combinatorial-auction} describes
the use of combinatorial auction for computing resource allocation in an EC market. The market includes 3
buyers, i.e., bidders, and one seller, i.e., the auctioneer. Buyers submit their bids to the seller that
specify the demands of computing resources. In particular, each bid indicates the demand of CPU resources and
energy resources. Then, the problem is to determine the winners and the price that each winner needs to pay.
To solve this problem, some optimization algorithms can be applied, e.g., dynamic programming (DP)~\cite{H.Zhang2017,Y.Hung2018,Y.Hung2020,S.Yang2020}, greedy algorithm~\cite{T.Bahreini2018,G.Gao2019,Y.Jiao2018} and graph neural networks~\cite{M.Prates2019}.

It is obvious that the combinatorial auction is appropriate to trade a bundle of complementary commodities and can effectively improve auction efficiency of allocating multiple commodities combination.
\subsubsection{Double Auction}
\begin{figure}[!t]
\begin{center}
\includegraphics[width=3.3in]{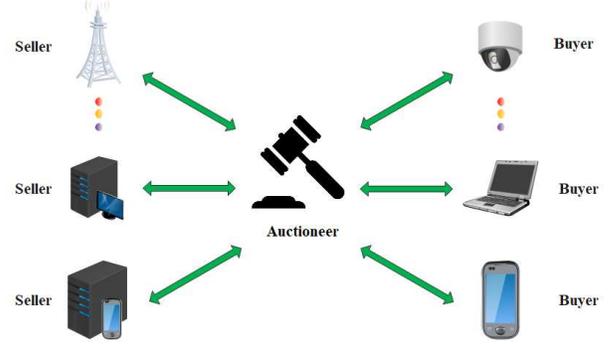}
\caption{The typical many-to-many structure of double auction.}
\label{fig:double-auction}
\end{center}
\end{figure}

A double auction~\cite{D.Friedman1993} is a multi-item auction that is widely applied to deal with optimal
allocation problems. Different from conventional auctions (e.g., English auction~\cite{V.Krishna2002}, Dutch
auction~\cite{V.Krishna2002}, first-price sealed-bid auction and second-price sealed-bid
auction~\cite{W.Vickrey1961}), double auction is not a one-to-many structure, but a many-to-many structure,
i.e., the number of sellers and buyers are both more than one, as depicted in Fig.~\ref{fig:double-auction}.
In the double auction, sellers and buyers respectively submit their asks and bids to an auctioneer, i.e., an
executor of the auction process. Then, the auctioneer sorts asks (asking prices) and bids (bidding prices) in
descending order and ascending order, respectively. After that, the auctioneer calculates the transaction
price $p^{*}$, i.e., a hammer price, by $p^{*}=(p^{b}_k+p^{a}_k)/2$, where $k$ denotes the largest index,
$p^{a}_k\leq p^{b}_k$, $p^{b}_k$ and $p^{a}_k$ denote the $k$th bidding price and asking price, respectively.
Finally, the winning buyer gets the resource and pays the corresponding seller $p^{*}$. The matching
relationship between remaining buyers and sellers and corresponding hammer prices can be determined by
repeating the above process.

In the next section, we will discuss how to adopt the combinatorial auction and double auction in the EC
market. In addition, we also consider other auction approaches that can be used for the resource allocation
in the EC market.

\section{Application of auction approaches in EC}
Considering the fact that different types of auction methods are suitable to address different types of
problems in EC. In this section, we present a comprehensive view on the applications of auction-based
approaches for EC under different types of auctions.

\subsection{Combinatorial Auction}
In the practical competitive market, the relationship between supplies and demands is always complicated. In
recent years, considerable attentions have been focused on combinatorial auction to deal with
combinatorial allocation
problems~\cite{L.Fawcett2016,T.Bahreini2018,A.Bandyopadhyay2020,F.Zhang2018,Y.Jiao2018,Y.Jiao2019,H.Zhang2017,Y.Hung2018,Y. Hung2020,P.Kayal2019,S.Yang2020,G.Gao2019}. As shown in Table \uppercase\expandafter{\romannumeral2}, we
summarize the above-mentioned work in terms of the issue, objective, market structure (i.e.,
seller, buyer, auctioneer, commodity), scenarios and advantages.

\begin{table*}
\centering
\scriptsize
\caption{Combinatorial Auction-Based Mechanism in EC}
\label{table2}
\begin{tabular}{|m{1.1cm}<{\centering}|m{1.3cm}<{\centering}|m{1.5cm}<{\centering}|m{1.1cm}<{\centering}|m{1.2cm}<{\centering}|m{1.1cm}<{\centering}|m{1.8cm}<{\centering}|m{1cm}<{\centering}|m{4cm}<{\centering}|}
\hline
\multirow{2}{*}{Ref.} &\multirow{2}{*}{Issue}&\multirow{2}{*}{Objective}& \multicolumn{4}{c|}{Market structure} & \multirow{2}{*}{Scenarios} & \multirow{2}{*}{Advantages}\\\cline{4-7} 
&\multirow{2}{*}{}&\multirow{2}{*}{}& Seller & Buyer &Auctioneer &Commodity&\multirow{2}{*}{}&\multirow{2}{*}{} \\
\hline \hline
\cite{L.Fawcett2016}&Resource allocation and pricing&Efficient allocation&SPs&Infrastructure providers&Orchestrator &Processor, memory, storage&FC&An popular auction-based resource allocation platform\\
\hline
\cite{T.Bahreini2018}&Resource allocation and pricing&Efficient allocation&Edge/cloud servers&Mobile users&Edge/Cloud server&VM instances(CPU, memory, storage)&Cloud/Edge computing&Guarantee IR and envy-free allocations, combining the advantages from position and combinatorial auctions, and greatly reduce the execution time\\
\hline
\cite{A.Bandyopadhyay2020}&Resource allocation and pricing&Higher income and allocation efficiency&Fog nodes&IoT user&Fog server&Fog node services(computation, storage, and networking related services)&FC&Guarantee TF, highly useful in various applications\\
\hline
\cite{G.Gao2019}&Resource allocation and pricing&Near-optimal social welfare&Edge cloud nodes&Mobile users&A construted platform&Virtual Machine (VM) resources&ECC&Guarantee IR, CE and TF, and the communication latency is greatly shorten\\
\hline
\cite{H.Zhang2017}&Computation offloading&QoS guarantee, and efficient computation offloading&MEC service providers&User equipments&MEC service providers&Wireless and computational resources&MEC&System performance of the proposed algorithm  outperform existing algorithms, consider demand heterogeneity of UEs\\
\hline
\cite{Y.Hung2018,Y.Hung2020}&Resource allocation and pricing&QoE guarantee, social welfare maximization&Base stations&Streamers&Edge system&Backhaul capacity and caching space&MEC&The proposed algorithm can be calculated in polynomial time, and greatly enhance overall system utility\\
\hline
\cite{S.Yang2020}&Task offloading&Efficient allocation, and task execution time minimization&MBS/SBS&Vehicles&MBS/SBS&Wireless and computing resources&MEC&Effectively reduce system overhead, and the average time for completing a task is minimized\\
\hline
\cite{F.Zhang2018}&Resource allocation and price&QoS guarantee, and the profit of fog nodes maximization&Fog service providers&Mobile users&A trusted third party&Computation resource (CPU and memory resource)&FC&Guarantee IR, CE, and TF, and one seller servers multiple buyers simultaneously\\
\hline
\cite{Y.Jiao2018,Y.Jiao2019}&Resource management and pricing&Social welfare maximization&ESP&Mobile users&ESP&Computation resource&MBN&Guaranteeing the TF, IR and CE\\
\hline
\cite{P.Kayal2019}&Energy consumption and communication costs&Energy consumption and communication costs maximization&Applications&End-user devices/fog nodes&Applications&Microservices&FC&The placement strategy outperform others, the network topology is robust and generates less energy overhead than other topologies, and effectively avoid leakage of private information or trading details\\
\hline
\end{tabular}
\end{table*}

In~\cite{L.Fawcett2016}, the authors constructed an auction-based resource allocation platform wherein the
resources as commodities mainly consist of processor, memory and storage. The platform includes three layers:
SPs, SRs and manager. The highest layer consists of many SRs, i.e., buyers, which need a bundle of resources
from the lowest layer, i.e, SPs (sellers). The auctioneer as the middle layer, which offers details of
available resources and hosts the auction processes. Then, a resource allocation function based on
combinatorial auction is applied to deal with the issue of the price competition between SPs and SRs. The
method aims to obtain the optimal price among SRs and guarantees to provide fairly recompense for SPs. In
addition, a detailed illustration about the allocation platform was given, which can better to illustrate the
core idea of the auction-based platform. However, this work did not verify the effectiveness of the proposed
method by any experiments. Moreover, the discussion for economic properties of the proposed mechanism was not
given.

It is challenging to obtain optimal revenue or social welfare
in MEC systems by effective allocation and pricing for limited edge/cloud resources. The authors
in~\cite{T.Bahreini2018} solved the problem of the virtual machine (VM) instances allocation and pricing between MDs
(buyers) and edge/cloud servers (sellers) by a combinatorial auction mechanism, called G-ERAP. The
G-ERAP is integrated with the combinatorial auction and the greedy algorithm~\cite{Y.Akcay2007}
to determine the winners and the payments. The novel G-ERAP combines the advantages from both
position auction and combinatorial auction for satisfying heterogeneous requests of mobile users (MUs).
The theoretical analysis demonstrated that G-ERAP can achieve IR while ensuring envy-free
allocation. Experimental results showed that the revenue and social welfare obtained by G-ERAP
are comparable to those obtained by CPLEX~\cite{IBM2009}. In addition, G-ERAP can greatly reduce
the execution time of the applications from MUs. However, as a two-level allocation mechanism,
G-ERAP arbitrarily offloads tasks to edge servers or cloud servers without any consideration for
different requirements and preferences of MUs. It is a promising research to consider the
preferences for MUs in designing allocation mechanisms where the preferences include the locations
of both MUs and SPs. Furthermore, the G-ERAP mechanism cannot ensure the property of TF.
It aims to achieve envy-free allocations by sacrificing TF.

To guarantee the TF, the authors in~\cite{A.Bandyopadhyay2020} proposed two types of truthful
mechanisms, i.e., the fixed price based fog node allocation mechanism (FixP-FogNA) and the combinatorial
auction based fog service allocation mechanism (CAuc-FogSA), both of which consider heterogeneous demands
from end users. FixP-FogNA and CAuc-FogSA aim to effectively allocate limited fog services (i.e.,
computation, storage, and networking related services) of fog nodes (sellers) to various users (buyers) by
adopting a fixed cost strategy and a combinatorial method, respectively. The winners and the payments are
obtained by adopting truthful approximation algorithm. It is noteworthy that the "fixed price" stands for the
fixed cost of corresponding fog node, namely a user should pay for using its service at a unit time slot in
FixP-FogNA. The CAuc-FogSA is an incentive compatible mechanism which can guarantee economic property of TF. Compared with FixP-FogNA, CAuc-FogSA
achieves higher revenue and allocation efficiency for fog service providers. On the contrary, FixP-FogNA has
less computational complexity than CAuc-FogSA. However, the validity and effectiveness of the proposed
mechanisms need to be demonstrated by experiments. Furthermore, the social welfare and other properties,
e.g., IR and BB, can be considered in the future work.

In contrast to~\cite{T.Bahreini2018} and~\cite{A.Bandyopadhyay2020}, authors in~\cite{G.Gao2019} addressed
the similar problem, i.e., allocating limited VM resource allocation between the MUs as buyers and
geo-distributed edge cloud nodes (ECNs) as sellers by a truthful auction-based VM resource allocation (AVA)
mechanism. The AVA mechanism includes two main algorithms, i.e., the greedy winning bid selection algorithm
and the payment determination algorithm, both of which can obtain near-optimal social welfare while
ensuring TF, IR, and computational efficiency (CE). The winners and the
payments are determined by the winning bid selection algorithm and payment determination algorithm
in~\cite{G.Gao2019}. Moreover, the communication latency can be greatly shorten by the AVA mechanism based on
the network paradigm of edge cloud computing (ECC)~\cite{U.Drolia2013}. The simulations supported by real
traces results demonstrate the validity and effectiveness of the AVA mechanism. However, the other factors,
e.g., bandwidth and quality of transmission, may need to be considered when evaluating the performance
of the proposed mechanism.

In fact, a reasonable and efficient resource allocation mechanism not only guarantees the economic
properties, e.g., IC, IR, and CE~\cite{T.Bahreini2018,A.Bandyopadhyay2020,G.Gao2019}, but also ensures the
QoE/QoS of UEs~\cite{Z.Ullah2020,R.Han2020,Y.Yino2020}.

\begin{figure}[!t]
\centering
\includegraphics[width=3.3in]{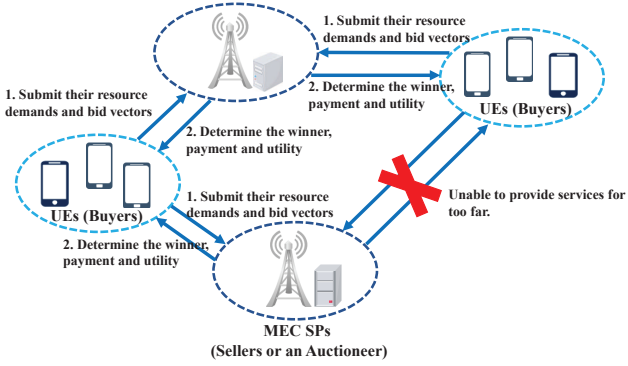}
\caption{The MEC network model.}
\label{fig:MEC-Network}
\end{figure}

In~\cite{H.Zhang2017}, the authors jointly considered economic properties and UEs' QoS to construct an
auction model based on matching relationship between multiple UEs and multiple MEC SPs in MEC networks. As
shown in Fig.~\ref{fig:MEC-Network}, UEs as buyers with heterogenous demands compete for limited wireless and
computational resources from the MEC SPs, while MEC SPs  play the roles of sellers and auctioneers. To solve
the above resource allocation issue, a multi-round-sealed sequential combinatorial auction (MSSCA) mechanism
is proposed, which consists of winners determination, bid strategy of users and the pricing process. The bid
strategy of users is inspired by multi-round priority rule. Then, the winner determination process is
transformed into a two-dimensional knapsack problem and can be solved by DP algorithms. The simulation
results demonstrated that MSSCA can improve the system performance while maintaining well QoS for
UEs. However, this work still has two shortages coming from the multi-round auction itself, i.e., the bidder
drop problem and resource waste problem~\cite{H.Zhang2017}, which may restrict further improvement of the
system performance. Moreover, it is more practical to assume that the valuation function is nonlinear with
the number of received resources~\cite{H.Zhang2017}, such as AR/VR~\cite{B.Krogfoss2020} and live video
streaming~\cite{Y.Hung2018}.

\begin{figure}[!t]
\centering
\includegraphics[width=2.5in]{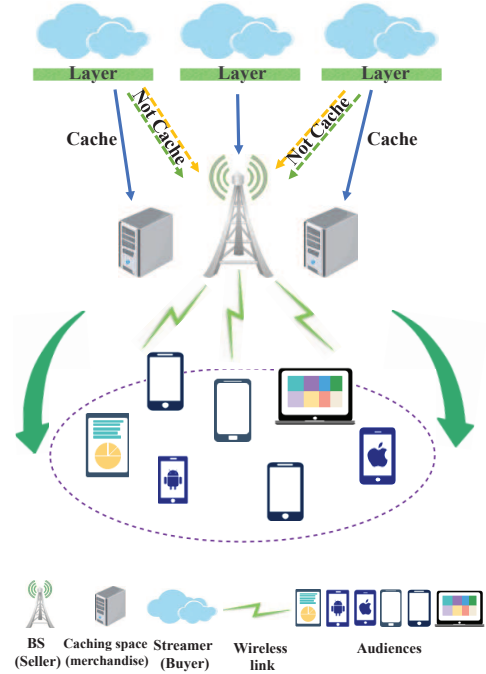}
\caption{System model of live video streaming services in MEC.}
\label{fig:LiveServiceForStream}
\end{figure}

The authors in~\cite{Y.Hung2018} tackled the limitations in~\cite{H.Zhang2017}. They considered that the
valuation function is given which is nonlinear with the number of received streamers. In order to improve the
QoE of live video streaming services~\cite{H.Schwarz2007,C.Zhan2018} under limited backhaul capacity and
caching space, they proposed an auction framework based on combinatorial clock auction (CCA) in streaming
(CCAS) framework~\cite{Y.Hung2018}. In such framework, base stations (BSs), streamers and caching space are
treated as sellers, buyers and merchandise, respectively (as shown in Fig.~\ref{fig:LiveServiceForStream}).
The edge system as an auctioneer hosts the auction process and determines the winners and caching space
allocation. The CCAS includes two stages: the clock stage and the supplementary phase. In the first stage,
sellers increase the prices and buyers submit their demands according to the prices in each round. In the
second stage, the auctioneer determines the winners and their payments by VCG mechanism. Then, the problem of
the caching space value evaluations and allocations was formulated as a variant of 0-1 knapsack
problem~\cite{G.B.Dantzig1957}, and was solved by DP algorithms in the CCAS. The mechanism can maximize
social welfare in polynomial time to the number of streamers. In order to further optimize the auction
process and reduce the complexity of calculation, a method of identifying equivalent package sets is applied.
The theoretical analysis demonstrates that CCAS can satisfy TF while achieving optimal efficiency.
The overall system utility can be significantly improved by CCAS which is demonstrated by the simulation
results. However, this work allocated the backhaul capacity by adopting a fixed rule when specifically
determining the caching space allocation~\cite{Y.Hung2020}. This may hinder its application and potentially
constraint system efficiency.

\begin{figure*}[!t]
\centering
\includegraphics[height=2.8in]{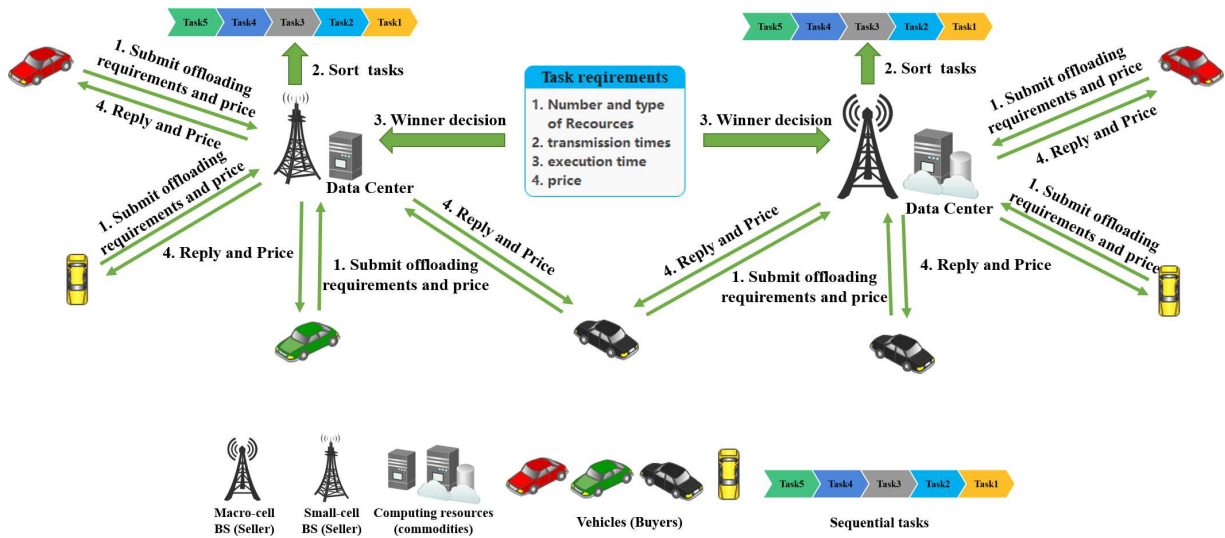}
\caption{The diagram of task offloading in MEC network.}
\label{fig:TaskOffloadSystem}
\end{figure*}

As an extension of~\cite{Y.Hung2018}, the authors in~\cite{Y.Hung2020} did more in-depth study for
the same problem by improving the proposed algorithm of~\cite{Y.Hung2018}. Two auction
frameworks, i.e., edge combinatorial clock auction (ECCA) and combinatorial clock auction in stream
(CCAS) are proposed in which streamers (buyers) are enable to bid for available backhaul capacity and caching
space from BSs (sellers) in the edge-aided cellular system. As shown in Fig.~\ref{fig:LiveServiceForStream},
the edge system acts as an auctioneer, who determines the backhaul capacity
and caching space allocation for streamers. To improve system efficiency and offer better QoS
services for audiences, ECCA and CCAS focus on the issues of the backhaul capacity and caching
space allocation (BCCSA), and the caching space value evaluations and allocations (CSVEA),
respectively. In ECCA, the problem of BCCSA is formulated as an optimization problem dominated by
the edge system, which specifically determines the allocation scheme of the backhaul capacity and
caching space according to the requests of streamers, and is solved by a three dimension DP
algorithm~\cite{Y.Hung2020}. However, the exponential relation between the number of streamers
and the minimum unit of capacity and space that converts to countable variables, which
restricts scalability of the algorithm. In CCAS, the problem of CSVEA is formulated as two
optimization problems dominated by the streamers and the edge system respectively, and is solved by
the DP algorithms. In CCAS, the backhaul capacity is evenly allocated by the edge system among the
streamers, and then streamers bid for the packet of caching space according to their requirements, which
can obtain maximum total valuation under limited caching space. The simulations showed that ECCA
outperforms the algorithms of CCAS, FSC, PC and FC~\cite{Y.Hung2018} in terms of social welfare and average
utility of each streamer. However, CCAS is a better method for achieving similar performance close to ECCA,
but with lower computation complexity and higher scalability. In fact, CCAS can achieve higher scalability
and lower computation complexity by sacrificing merely a little efficiency. To sum up, the property of
TF and the optimality of efficiency are both guaranteed by ECCA and CCAS, as demonstrated
theoretically. However, ECCA and CCAS did not take into account the latency requirements in the
utility functions.

\begin{figure}[!t]
\centering
\includegraphics[width=3.3in]{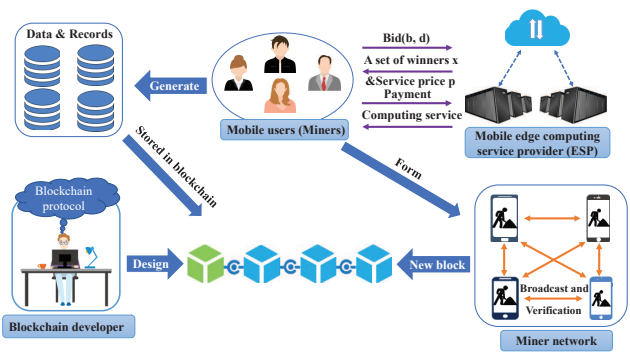}
\caption{Resource allocation for mobile blockchain in EC system.}
\label{fig:BlockchainNetwork}
\end{figure}

In order to improve task offloading efficiency in terms of tasks execution time, the authors
in~\cite{S.Yang2020} described the task offloading problem as a multi-round sequential combination
auction by considering mobile vehicles and macro-cell/small-cell base station (MBS/SBS) to be the buyers
and sellers, respectively. The mobile devices have heterogeneous resource requirements and the MBS/SBS
deploy various service nodes with limited wireless and computing resources. As shown in
Fig.~\ref{fig:TaskOffloadSystem}, many vehicles with offloading tasks submit diverse requirements
and bid prices for nearby service nodes in sequence. After receiving the requests from the vehicles, the
service node applies the winner decision algorithm to determine the winning vehicle and the
payment. The optimal matching relationship between vehicles and service nodes can be transformed into
a multi-dimensional grouping knapsack problem, and then is addressed by a DP algorithm. Compared to the
existing task offloading algorithms~\cite{M.Chen2016,T.Zhao2017,T.Z.Oo2017}, the proposed algorithm
can obtain shortest average task competition time, effectively reduce system overhead, and is
demonstrated by simulation results. However, a more reasonable and significant factor, i.e., preferences
of both vehicles and service nodes, can be considered to be an extension. Furthermore, the economic
properties, e.g., TF and IR, should be guaranteed in the proposed mechanism.

Similar to~\cite{H.Zhang2017,Y.Hung2018,Y.Hung2020,S.Yang2020}, the authors in~\cite{F.Zhang2018} took
economic properties, users' QoS and resource allocation efficiency into account to design a combinatorial
auction mechanism based on a dynamic resource allocation model. The model is composed of fog nodes, i.e.,
sellers, each of which owns heterogeneous computing resources (contains CPU and memory resource) and is
deployed around buyers, MDs, i.e., buyers, which need to offload heterogeneous tasks to fog nodes, and a
trustworthy third party, i.e., an auctioneer, which administrates the auction process. The winner and the
price are determined by the winner determination rule and the pricing rule, respectively. In addition, the
pricing model has three billing methods: on-demand, daily and auction billing, which is defined according to
the tasks be processed. They divide the tasks in terms of execution time, delay-sensitive and computational
complexity. Moreover, unlike McAfee's mechanism~\cite{R.P.McAfee1993} that one seller can only serve one
buyer, the proposed mechanism allows one seller to serve multiple buyers simultaneously. Particularly, the
resource overbooking and prediction algorithms based on LSTM-enabled~\cite{S.Hochreiter1997} neural network
and service level agreement violation (SLAV) feedbacks are proposed, which can dramatically eliminate the
impact of low resource utilization even if during non-peak-hour. Thus, the mechanism achieves a high degree
of QoS satisfaction for diverse tasks. The experiment results demonstrated that the proposed mechanism
guarantees TF, IR, BB and EE. It is obvious that the proposed mechanism
is superior to McAfee's mechanism. Moreover, similar to~\cite{A.Bandyopadhyay2020}, the proposed
mechanism can maximize the profit of SPs, i.e., fog nodes. However, the effectiveness of task scheduling
was not considered in this work.

As an emerging decentralized technique, blockchain has been gained considerable attentions. Recently, blockchain technique has been widely used in security and reliability field~\cite{T.Salman2019,A.Asheralieva2020}. Unfortunately,
mining process cannot be supported by MDs with limited computing and storage capacities which hinders
applications and developments of blockchain technique in EC.

In mobile blockchain networks (MBNs), the limited computing power and storage space of MDs (miners) lead to
the mining issues. To address the issue, the authors in~\cite{Y.Jiao2018} transformed the mining task
offloading problem as an auction by treating the edge computing service provider (ESP) plays the roles of the
seller and the auctioneer, and MDs as buyers. As shown in Fig.~\ref{fig:BlockchainNetwork}, the ESP owns edge
computing servers (ECSs) and deploys them around MDs. The MDs compete for computing services of nearby ECSs
to process their mining tasks, and ESP decide the winners and the payments that MDs should pay ECSs.
Then, a auction-based mechanism for resource allocation in MBNs was proposed to achieve social welfare
maximization while satisfying the desirable economic properties, i.e., IR and TF.
In addition, the mechanism contains an algorithm which is integrated with the greedy
algorithm~\cite{Y.Akcay2007} and the VCG auction~\cite{E.H.Clarke1971,T.Groves1973,L.M.Ausubel2006}
that can obtain the winners and the payments while enhancing the trading frequency of participants in
the system. The simulations demonstrated that the proposed mechanism can obtain optimal allocation for
computing resources. However, it is not practical to only consider MDs (miners) with constant demand.
Moreover, the network effects function need to be verified by experiments.

Unlike~\cite{Y.Jiao2018} in which the buyers have constant demands, the authors in~\cite{Y.Jiao2019}
considered the buyers with multi-demand in constructing auction-based mechanism. Then, similar
to~\cite{Y.Jiao2018}, a combinatorial auction-based allocation mechanism for computing resource was
introduced, where the cloud/fog computing service provider (CFP) acts as the auctioneer and the
seller, and MDs (miners) act as buyers. The MDs compete for nearby cloud/fog servers (CFSs) to
process the mining tasks, where CFSs are deployed and managed by the CFP. The allocative
externalities was considered in designing auction mechanism due to the competition among
miners. Note particularly the two bidding schemes of~\cite{Y.Jiao2019}, i.e., the constant-demand
scheme and the multi-demand scheme, that both schemes can obtain social welfare maximization meanwhile
ensuring desired economic properties (e.g., TF and IR) and CE. The main difference between the two bidding
scheme is the restrictions on bidding. More concretely, each miner of the constant-demand scheme can only bid
for resources in fixed quantity. In contrast, each miner of the multi-demand scheme can bid for resources in
diverse quantity and submit their interesting demands and bids. The real-world experiments demonstrated that
the proposed mechanisms can achieve optimal resource allocation, and offer valuable guidance
for the application of blockchain technique in EC. However, enhancing total revenue
of the server provider can be investigated in the future work~\cite{Y.Jiao2018,Y.Jiao2019}.

Considering undesirable characteristics of MDs in EC, such as short-term power supply, which
highlights the significance of energy overhead. To minimizes energy overhead and communication costs, the
authors in~\cite{P.Kayal2019} proposed a distributed fog service placement (DFSP) algorithm based on
an iterative combinatorial auction. The auction process contains the interactions between fog
nodes (buyers) and applications (sellers), where sellers own a set of microservices and
computing resources. In each iterative of an auction, fog nodes bid for a bundle of microservices
(commodities) of the designated application, where the designated application is determined by fog
nodes before bidding. To determine which fog nodes obtain the bundle of microservices from the
designated application and the price of the microservices, a dynamic pricing scheme is implemented.
Unlike conventional auction model, their model is fully distributed, i.e., without a central
auctioneer, and buyers decide whether or not to send their private data and to whom. This
decentralized framework can avoid leakage of private information and trading details. The numerical
examples showed that DFSP achieves minimum total cost of existing algorithms~\cite{C.Pham2017,A.Beloglazov2012}.
Then, they evaluate the quality of the placement strategy of DFSP in terms of the CPU utilization,
communication cost and the number of fog nodes, the results well verified its performance outperforms the other
algorithms~\cite{C.Pham2017,A.Beloglazov2012}. Furthermore, the linear network topology of proposed
system is more robust and generates less energy overhead than previous
topologies, which was proofed by some experiments~\cite{C.Pham2017}. However, the proposed mechanism
only allows an application trading with a fog node.

In summary, in this section, we have reviewed the existing studies of resource allocation, pricing scheme,
task offloading, optimize energy consumption, etc. in EC. Compared to existing works that without
combining with combinatorial auction methods, the effectiveness of those combinatorial
auction-based methods are theoretically and experimentally demonstrated. However, in multi-user and
multi-server scenarios, it is still a challenge to construct an effective and truthful auction
mechanism, manage matching relationship between UEs with personalized requirements and SPs with
heterogeneous resource distribution, and achieve optimal allocation scheme while ensuring that the system
satisfies all the desired economic properties. Furthermore, some authors realize the essentiality
and necessity of privacy preserving and decentralized mechanism in EC, and the trend is continuing
for a long time to come.

\subsection{Double Auction}
A double auction~\cite{D.Friedman1993} is a popular method for its typical many-to-many structure and
is widely applied in real-world markets~\cite{X.Zhai2018,L.Lu2018,S.Zhan2016}. Recently, the number of
IoT MDs increases tremendously, and it has seen various applications and research of double auction-based
algorithms in multi-user and multi-server scenarios of EC.

In~\cite{A.Jin2018}, the authors addressed the issue of computing resource sharing via designing an
incentive-compatible auction periodical mechanism (ICAM) between MDs as service users (buyers) and
cloudlets as SPs (sellers). ICAM consists of three phases, i.e., winning candidate determination,
assignment$\&$pricing, and winner elimination. In the first phase, the auctioneer determines the
winning candidates according to an ascending buyer set and a descending seller set. In the second phase,
a pricing rule is adopted to ensure the TF for buyers. In the last phase, surplus winners are
abandoned and one buyer can only match one seller. The winers and their payments are determined by
Algorithms: ICAM-A$\&$P and CAM-WE in~\cite{A.Jin2018}. The numerical results showed that ICAM can enhance
resource utilization of cloudlets while guaranteeing TF, IR, and BB. Moreover, ICAM can
attain around 50\% of the system efficiency for the allocation scheme. However, ICAM only allows to match
a seller in each auction. It is impractical and may not fully utilize the ability of resource-rich SPs, which
could have offered services for multiple users. Furthermore, they did not consider crucial influence of the
locality characteristics of SPs on system efficiency.

Different from the matching manner in~\cite{A.Jin2018}, some more reasonable and comprehensive mechanisms
were proposed in the literature, i.e.,~\cite{Y.Yue2018,Y.Yue2019,W.Sun2018,R.Zhang2019,H.Hong2020}. In
particular, the authors in~\cite{Y.Yue2018} eliminated the shortcoming of the matching manner
in~\cite{A.Jin2018}, and jointly considered scarce computing resources and location properties. They
introduced a single-round double auction mechanism based on breakeven (SDAB) in MEC. In the auction-based
model, MDs (buyers) with individual awareness and preferences, compete for edge servers (sellers) with
limited computation resources. As a trustworthy administrator, the auctioneer manages the whole auction
process. The finial winners and payments are determined by the SDAB algorithm. Compared with ICAM, SDAB
allows one seller to support multiple buyers with offloading demands. Moreover, network economics and
resource allocation were both considered in SDAB to maximize the number of resource tradings. The theoretical
analysis and the numerical results verified that SDAB can guarantee IR and TF. The system efficiency of SDAB
outperforms ICAM. It was shown that the number of successful trades obtained by SDAB is higher than those of
ICAM by around 20\% when the number of buyers increases to 50 under the unchanged number of sellers. However,
SDAB cannot guarantee BB, and only considered the homogeneous tasks of MDs.

\begin{figure}[!t]
\centering
\includegraphics[width=2.8in]{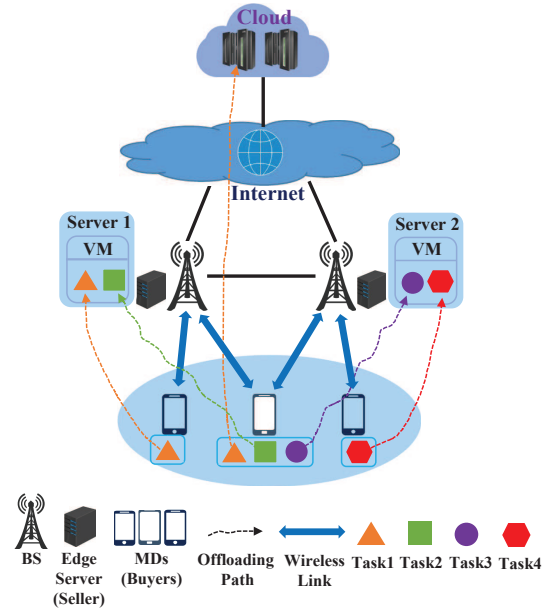}
\caption{A multi-task cross-server architecture in MEC.}
\label{fig:MultiTaskModel}
\end{figure}

Considering BB and heterogeneity of both resources and tasks, the authors in~\cite{Y.Yue2019}
designed a cross-server resource allocation algorithm based on double auction (MADA) to enhance the system
efficiency in MEC. As shown in Fig.~\ref{fig:MultiTaskModel}, a supply-demand relation between MDs (buyers)
and edge servers (sellers) is modeled as an auction where MDs with heterogeneous tasks bid for diverse
applications (commodities) assigned in VMs of edge servers. In the model, the closest edge server as an
auctioneer hosts the auction market. MADA is composed of three stages, i.e., remote cloud assistance, natural
ordering and price $\&$ winner decision. In the first stage, the auctioneer decides whether to offload tasks
to remote cloud according to the delay tolerance. In the second stage, they auction all types of applications
(apps). In the last stage, the prices and winners are determined by price$\&$winner algorithm. In particular,
MADA can be calculated in polynomial time. The simulation results showed that MADA can effectively allocate
resources while satisfying TF, IR and weakly BB. In addition, MADA can obtain higher system efficiency than
the existing studies~\cite{A.Jin2018,Y.Yue2018}. Under the fixed number of edge servers, the overall utility
of MADA is always higher than that of ICAM~\cite{A.Jin2018} via regulating number of buyers. The number of
successful trading of ICAM is lower than MADA by around 50\% when the number of buyers meets 50. However, the
large growth of the number of both buyers and sellers will lead to higher computational complexity.
Furthermore, this work cannot obtain the optimal social welfare.

With rapid expansion of application for industrial Internet of Things (IIoT), the traditional
resource allocation strategies are unable to satisfy the growing various requirements for edge
devices. Thus, it is crucial to enhance resource utilization rate for an MEC-based IIoT scenario.
Considering both network economics and locality constraints, the authors in~\cite{W.Sun2018} described
the interaction between IIoT MDs and edge servers as a double auction, and design two
double auction algorithms, i.e., a breakeven-based double auction (BDA) and a dynamic pricing based
double auction (DPDA) to enhance the efficiency of resource utilization in MEC. The model is composed
of a set of lightweight edge servers, i.e., sellers, each of which is equipped with limited computing
capability, a number of IIoT MDs, i.e., buyers, which need computing resources (commodities) to
process computing-intensive tasks, the SP, i.e., an auctioneer, which manages the auction and
determines matching pairs and the price. BDA and DPDA both utilize dynamic pricing strategy to
allocate resources and achieve the number of successful matching pairs maximization, while
satisfying the economic requests for MDs. The theoretical analysis shows that both algorithms can
guarantee the economic properties of TF, IR, BB and
EE. As shown in the simulation results, the utility of edge servers and IIoT MDs in
DPDA were both higher than that in BDA under different number of IIoT MDs. Similar to
ICAM~\cite{A.Jin2018}, BDA maintains the TF between buyers and sellers by applying a
breakeven, and DPDA achieves higher system efficiency by sacrificing TF. Moreover, the
number of successful trades of DPDA is the highest compared to BDA and ICAM under different number
of IIoT MDs. In summary, BDA and DPDA can substantially improve the system efficiency for MEC-based
IIoT system. Although the performance of DPDA outperforms BDA and ICAM, DPDA cannot guarantee
TF of sellers.

The above existing works~\cite{A.Jin2018,Y.Yue2018,Y.Yue2019,W.Sun2018} improves the resource
allocation efficiency mainly focus on the objective level, they ignore the subjectivity of
participants in practical scenarios. To further improve resource utilization rate and achieve
higher social welfare, the authors in~\cite{R.Zhang2019} addressed the issue of computing resource
allocation while considering the preference of MDs. In the D2D-assisted MEC model, a set of
MDs have heterogeneous computing requests act as buyers and sellers, the edge server with limited
computing resources (commodities) play the roles of the seller and the auctioneer. Compared with the
existing model~\cite{A.Jin2018,Y.Yue2018,Y.Yue2019,W.Sun2018}, the model of~\cite{R.Zhang2019} can
fully utilize computing resources of idle MDs. Thus, a practical and comprehensive auction scheme
for computing resource allocation (ASCRA) mechanism in MEC system is proposed. The ASCRA is composed
of three stages, namely, identification confirmation, candidate selection, and matching$\&$pricing.
In the first stage, ASCRA-IC algorithm identifies each device as a seller or a buyer based
on their remaining resources. In the second stage, the resource condition and the delay condition
are both considered in ASCRA-CS algorithm to select the candidate pairs among buyers and sellers. In
the last stage, ASCRA-MP algorithm determines the winners and payments. The numerical results
showed that the performance of ASCRA outperforms ICAM~\cite{A.Jin2018}, BDA and DPDA~\cite{W.Sun2018}.
Compared to BDA, DPDA, and ICAM, ASCRA has the highest number of successful trades under different
number of sellers, which due to idle MDs can share and sell their idle computing
resources. The least number of successful trades is ICAM, which due to the usage of breakeven.
Importantly, ASCRA meets all the desired economic properties. However, similar to DPDA, ASCRA aims
to realize more higher system efficiency by sacrificing TF of sellers.

To achieve efficient resource allocation, the number of successful trades as a common objective is selected
by~\cite{A.Jin2018,Y.Yue2018,Y.Yue2019,W.Sun2018}. Different from those works, the authors
in~\cite{Z.Li.Z2019} enhanced the efficiency of computing resource trading in edge-assisted blockchain-based
IoT. A trusted resource-coin loan system based on credit is established which consists of edge servers, i.e.,
lenders, which act as SPs own surplus resource coins, IoT devices, which play different roles include
resource-coins lenders, borrowers and idle IoT devices, and a broker, i.e., the manager, which manages the
trading market and provides trading-related services for participants. To meet the requirements of fast
payment and frequent trading, a credit-based payment scheme is presented. Based on above, an iterative
double-auction algorithm is executed by the broker to solve the optimization problems of resource-coin loan
and loan pricing. Moreover, the broker performs the rule of loan pricing not only can extract the hidden
information of participants, but also incentivize them to willing to trade. The theoretical analysis and
simulations demonstrated that the proposed algorithm can achieve social welfare maximization while satisfying
truthfulness, IR and BB. In addition, the algorithm can also protect privacy of trading participants.

Similar to the work in~\cite{Z.Li.Z2019} that extracted hidden trading information to achieve optimal
resource allocation by frequent iteration, the authors in~\cite{Z.Li2019} designed an optimal
iterative double auction-based algorithm for computing resource trading in blockchain network which
achieves social welfare maximization while preventing participants from privacy leakage. They
constructed a pure peer-to-peer (P2P) trading system for computing resources based on blockchain
technique, which can ensure security and TF for each resource trade. The system is
composed of the edge-CC service providers, i.e., sellers, which own redundant
computing resources and sell them to IoT devices or nearby edge nodes, IoT devices, which
play the roles of buyers, sellers and idle nodes, a broker, i.e., the controller,
which adjusts and manages the trading market via a smart contract. The algorithm alternatively
optimizes BAP, BMP and SMP optimization problems while satisfying the price rules to obtain optimal
prices and winners by multiple iterations. Both the theoretical analysis and experimental results
demonstrated that the proposed algorithm satisfies incentive compatibility (IC), IR,
and BB. Unlike ICAM~\cite{A.Jin2018}, SDAB~\cite{Y.Yue2018} and BDA~\cite{W.Sun2018}
utilize the breakeven approach to guarantee TF of both buyers and sellers, the proposed
algorithm adopts the effective price rules to achieve truthful trading while motivating both buyers
and sellers participate in resource trading. However, this work tackled the issue of resource
allocation supported by a centralized-control framework, which always meets a performance
bottleneck when numerous deals occur.

Considering the emerging challenges of the performance bottleneck caused by the centralized
framework~\cite{A.Jin2018,Y.Yue2018,Y.Yue2019,W.Sun2018,R.Zhang2019,Z.Li.Z2019,Z.Li2019}, the
authors in~\cite{A.Zavodovski2019} proposed a secure and incentive-compatible double auction
mechanism for computational resource sharing based on a decentralized EC framework, called
DeCloud. DeCloud includes three major types of participants, i.e., user clients, act as buyers
which need computational resources, service providers, play the role of sellers which provide or
share limited resources for buyers, and a distributed ledger, acts as a trusted execution
environment which supports and executes auction algorithms based on smart contracts. In DeCloud,
the impact of both potentially malicious providers and clients can be eliminated by distributed
ledger backed by blockchain technique. Hence, DeCloud can prevent privacy leakage and information tampering of clients.
\begin{figure}[!t]
\centering
\includegraphics[width=3.3in]{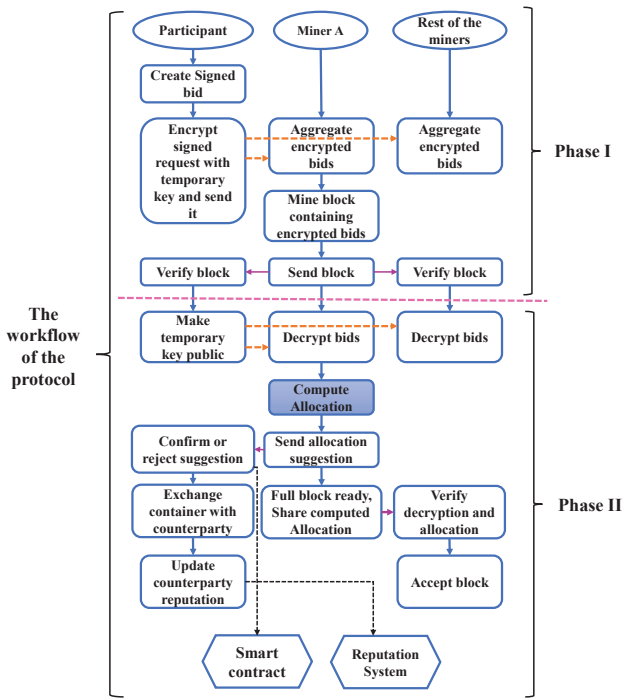}
\caption{The workflow of the two-phase bid exposure protocol.}
\label{fig:TwoPhaseProtocol}
\end{figure}
As shown in Fig.~\ref{fig:TwoPhaseProtocol}, DeCloud is integrated with a truthful bid expose
protocol includes two phases, i.e., sealed bids phase and allocation$\&$agreement phase. It is
noteworthy that the protocol aims to maintain sealed bids based on a transparent distributed
ledger. To achieve high-quality and flexible matching, a heuristic matching mechanism is designed
by utilizing an extensible bidding language with strong expressive power. Based on above, DeCloud
can tackle well the relationship between supply and demand with a high degree of heterogeneity.
Then, the optimal prices and the winners can be obtained by a DP algorithm in polynomial time.
The economic properties of IC, IR and strongly BB were proved to be guaranteed by the theoretical analysis.
In addition, the experiments were conducted on Google cluster-usage
data showed that DeCloud can enhance optimal welfare from 70\% to over 85\% according to the exact
market conditions. Nevertheless, this work cannot provide resource allocation services in the multi-task
cross-server scenario.

In~\cite{W.Sun2020}, the authors considered the issue of multi-task cross-server resource allocation
while considering profit-driven nature of participants in blockchain-based MEC. A DPoS-based
blockchain technique was adopted to achieve a decentralized and tamper-proof resource allocation
consensus mechanism, which can protect users information from tampering by malicious edge servers.
Then, the two-sided matching relation between MDs and edge servers is modeled as a double auction,
where MDs, edge servers, and the computing and storage resources of edge servers are treated as buyers, sellers and commodities, respectively. Then, two double auction mechanisms are designed, i.e., a
double auction mechanism based on breakeven (DAMB) and a breakeven-free double auction mechanism
(BFDA), where BFDA is more efficient than DAMB, and both of them satisfy TF, IR and BB.
Particularly, BFDA can address more tasks by sacrificing TF of buyers, which is similar
as DPDA~\cite{W.Sun2018} and ASCRA~\cite{R.Zhang2019}. Furthermore, the time complexity of both DAMB
and BFDA is low. In order to ensure untampered and secure resource allocation, like DeCloud~\cite{A.Zavodovski2019},
the proposed algorithms are executed as smart contracts based on the DPoS-based mechanism in blockchain.
The simulations showed that the performance of both DAMB and BFDA outperforms EDA and TIM~\cite{A.Jin2016}
in terms of execution time and system efficiency. BFDA has better performance than DAMB in terms
of average utilization rate of applications and average offloading rate of tasks. However, a more
practical model with high mobility and time-varying of MDs for resource allocation can be investigated
in the future work.

To jointly address the high-speed mobility and time-varying of moving vehicles in resource
allocation, the authors in~\cite{K.Xiao2020} proposed a dynamic allocation algorithm of edge
resources (DAER) based on the double auction mechanism. In order to prevent third parties from
tampering with users information, a resource transaction architecture backed by blockchain is
constructed which acts as an auctioneer in an auction. In DAER, user vehicles act as buyers compete
for resources from SPs, i.e., sellers, which own limited computing resources and capacities
(commodities). The DAER algorithm is performed by a smart contract supported by a blockchain architecture,
which jointly considers personalized demands of both vehicles and SPs. More importantly, the
mobility of vehicles is considered in DAER, where the mobility leads to the location and the
receiving service area of the moving vehicle are changing continuously. In order to provide resource
freezing and pre-allocation service for moving vehicles, and maximize satisfaction between vehicles
and SPs, three key algorithms, i.e., the state search algorithm, the selection of resource blocks
and SPs algorithm, and resource freezing and pre-allocation algorithm, are designed. The first
algorithm based on state prediction model can predict the next destination of the moving vehicle.
The second algorithm can significantly decrease the service resources waste. The last
algorithm can obtain an optimal freezing and pre-allocation strategy for the driving system to achieve
the overall satisfaction value maximization of vehicle and SPs. In fact, the last algorithm is an improved
genetic algorithm which determines the winner and the payment. The experiments showed that the
performance of DAER outperforms some existing studies~\cite{M.He2004,W.Zhang2013,S.Barbarossa2014,J.Feng2017}
in terms of task processing capabilities, resource utilization rate, and total satisfaction value. However,
this work did not consider the issues of road congestion.

The aforementioned approaches only allow one task of an MD to be assigned to one SP, which potentially
reduces resource utilization especially for those resource-rich SPs and against shorten the execution time of
tasks. To overcome the challenge, the authors in~\cite{H.Hong2020} proposed a double auction-based mechanism
for task offloading which supports one task of an MD can be serviced by multiple SPs. In the mechanism, an
auction model is introduced to describe the service relationship between MDs and edge servers, where MDs act
as buyers, and edge servers, i.e., seller, which have limited computational resources (commodities).
Importantly, the novelty of the model is that one task of an MD can be split into
multiple subtasks and distributed among multiple edge servers. In essence this means that all subtasks may be
processed parallel over different edge servers. Obviously, it is quite effective in reducing tasks execution
time for task offloading system. In addition, the edge servers own diverse service prices and qualities. To
efficiently offload heterogeneous tasks and obtain optimal social welfare, the winning bids determination
problem is transformed into a cost flow minimization problem. Then, the cost-scaling push-relabel
algorithm~\cite{A.V.Goldberg1997} is applied to solve the problem in
polynomial time. The proposed mechanism satisfies IR, strong BB and CE. Compared to DPDA~\cite{W.Sun2018},
the proposed mechanism can obtain higher social welfare and better scalability. However, unlike the previous
mechanism, such as ICAM~\cite{A.Jin2018}, SDAB~\cite{Y.Yue2018} and BDA~\cite{W.Sun2018}, this work cannot
utilize the breakeven approach to achieve TF for both buyers and sellers.

In recently years, we have witnessed the widely applications of reinforcement learning (RL) in solving the problems of decision-making strategy~\cite{S.Wang2018,D.Zhao2020,M.H.Ling2019}. It is enlightening to utilize the popular auction approaches that combined with RL to address the resource allocation problem.

In~\cite{Q.Li2020}, the authors formulated the resource allocation problem as a double auction by seeing MEC
servers, UEs/IoT devices, and a broker as the resource sellers, buyers and an auctioneer, respectively. In
order to obtain the Nash equilibrium, the experience-weighted attraction (EWA)~\cite{V.L.Smith1962} algorithm
is designed and running on all auction participants, where the EWA algorithm is integrated with RL and belief
learning~\cite{C.Camerer1997}. Moreover, the participants with limited resource can support EWA algorithm due
to the low computational complexity of it. Thus, buyers and sellers can adapt to various changes and
dynamically change their asking and bidding strategies based on the EWA algorithm in the auction market. The
experiment results showed that the EWA algorithm can obtain an outstanding convergence performance while
meeting profit demands of buyers and sellers, respectively. Specifically, the sellers will raise their asking
prices with the increase of average cost but cannot raise extremely due to the constraints of the broker. On
the contrary, the sellers will reduce their asking prices with the increase for the higher average
capacity of their servers. This means that sellers can service more buyers and achieve higher profit.
Moreover, the experiments showed that the buyers will not raise the bidding price a lot when
increase their average value due to the number of buyers is large. However, the economic properties of the
proposed mechanism need to be demonstrated theoretically and experimentally.

In order to allocate limited resources efficiently and reasonably, the above works design
auction-based mechanism almost focus on the related attributes of price. However, the nature of EC
system indicates that fully utilize price and various nonprice attributes is more effective in
resource allocation meanwhile ensuring the network economics.

\begin{figure}[!t]
\centering
\includegraphics[width=3.3in]{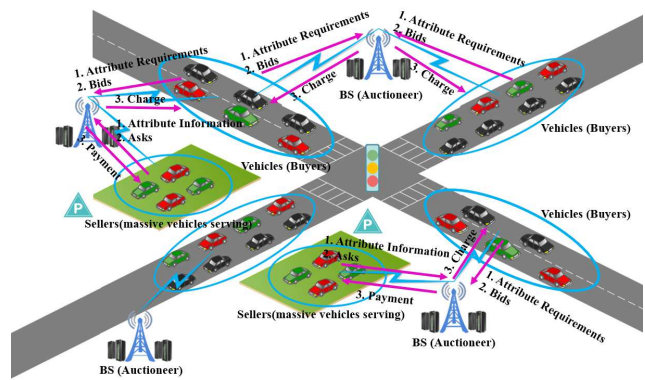}
\caption{System architecture of VFC.}
\label{fig:VFC-System}
\end{figure}

In~\cite{X.Peng2020}, the authors jointly considered the price and nonprice attributes (e.g.,
location, reputation and computation capability) to design a multiattribute-based double auction (MADA)
mechanism in vehicular fog computing (VFC). As shown in Fig.~\ref{fig:VFC-System}, auction-based
VFC system consists of some BSs and massive vehicles services, i.e., sellers, which provide various
services for vehicles in the vicinity, and client vehicles, i.e., buyers, which need computing and
storage resources (commodities) to process latency-sensitive tasks. Considering both the limited
resources and transmission distance, the VFC system is divided into some subsystems based on the
coverage of a BS. Then, the BS acts as an auctioneer to decide the winner and price. The MADA
includes three main phases: matching, assignment and winner determination and pricing. In the first and
second phase, the Kuhn-Munkres (KM) algorithm~\cite{H.W.Kuhn1955,J.Munkres1957} is applied to tackle the
issue of maximizing weighted matching while maximizing resource utilization. In the last phase, MADA
adopts a truthful and computationally efficient pricing algorithm~\cite{D.Yang2011} to determine the
winner and the price. Theoretical analysis and experimental results demonstrated that the MADA can guarantee
TF, IR, BB and CE. Furthermore, MADA can be calculated in polynomial time. However, considering
individual awareness and behavioral preferences of both buyers and sellers is a promising research direction
to improve allocation efficiency.

Blockchain technique can achieve decentralized, tamper-proof and security resource allocation consensus
mechanism, and has been widely applied to various field, such as financial transactions~\cite{C.H.Liao2020},
data storage~\cite{W.Liang2020}, IoT~\cite{C.K.Pyoung2020}, healthcare~\cite{S.WangJ2018}, etc. However,
limited communication and computing capabilities of MDs cannot meet the high computing requirements for
blockchain mining process, which restricts the applications and developments of the blockchain technique in EC.

To overcome the above problem, the authors in~\cite{X.Liu2019} introduced a combinatorial double
auction-based mechanism for the VM instances (commodities) trading, where MDs (buyers) act as
miners compete with each other to migrate mining tasks to edge server providers (sellers). Each ESP owns
different types of VM instances, varying in quality and price, and each MD need a package of different types
of VM instances to process mining tasks. Two allocation algorithms, i.e., a step greedy algorithm (STGA) and
a smooth greedy algorithm (SMGA) are proposed to determine the winners and allocate resources. In addition, a
payment scheme algorithm (PSA) based on VCG auction is proposed to calculate trade prices, and a group buying
rule is adopted to enhance the total utility of the system. The proposed mechanism can guarantee the economic
properties of TF, IR and BB. In addition, its computational efficiency is high. The experimental results
demonstrated that the performance of STGA outperforms SMGA, TACD~\cite{G.Zhou2017}, and
CDARA~\cite{P.Samimi2016} in terms of total utility, percentage of winning miners, and utilization of edge
servers. However, the overall time complexity is increasing tremendously with the increase of the number of
miners. Moreover, the available resources from nearby non-mining-devices need to be fully utilized.

Considering the available resource from idle devices in the vicinity, the authors in~\cite{S.Guo2020}
dealt with the mining issues by a double auction mechanism based on a constructed collaborative mining
network(CMN), where CMN is composed of numerous non-mining-devices and edge cloud. Similar to the
model in~\cite{X.Liu2019}, MDs (buyers) act as miners need to offload mining tasks to sharing-devices
or edge cloud (seller). Therefore, the mechanism provides two kinds of offloading objects, i.e., non-mining
sharing-devices and edge cloud. The task offloading problem is converted to a double auction game
when miners offload mining tasks to neighboring non-mining-devices within the CMN. Then, the optimal
auction price can be found by analyzing the Bayes-Nash Equilibrium~\cite{S.Zou2018}. When the offloading
object is edge cloud, the interaction between edge cloud operator (ECO) and CMNs can be well modeled as a
price-based optimization problem by stackelberg game. Then, the optimal price of ECO and the maximal profit
of CMN can be obtained by NE analysis. The simulation results showed that the profits of CMNs under the
proposed mechanism are higher than PECRM~\cite{Z.Xiong2018} by 6.86\% on average. However, this work
did not discuss some economic properties (e.g., IR and BB) and incentives for sharing-devices.

In this section, we review the existing literature of double auction-based resource allocation
approaches in EC. We summarize those works along with references in
Table~\uppercase\expandafter{\romannumeral3}. As shown in the table, mostly works focus on
resource allocation and pricing in various practical scenarios, such as MCC, MEC, IoT, IIoT, VFC,
blockchain-based IoT, blockchain-based IIoV, etc and aim to maximize the system efficiency and
social welfare of those systems while satisfying the desired economic properties of TF,
IR, BB, etc. Moreover, some works combine blockchain technique
with the existing auction-based mechanism to prevent privacy leakage and tampering. In the next
section, we will discuss the use of multi-round or online auction approaches in EC.

\begin{table*}
\centering
\scriptsize
\caption{Duble Auction-Based Mechanism in EC}
\label{table3}
\begin{tabular}{|m{0.8cm}<{\centering}|m{1.3cm}<{\centering}|m{1.5cm}<{\centering}|m{1.1cm}<{\centering}|m{1cm}<{\centering}|m{1.3cm}<{\centering}|m{1.8cm}<{\centering}|m{1.3cm}<{\centering}|m{4cm}<{\centering}|}
\hline
\multirow{2}{*}{Ref.} &\multirow{2}{*}{Issue}&\multirow{2}{*}{Objective}& \multicolumn{4}{c|}{Market structure} & \multirow{2}{*}{Scenarios} & \multirow{2}{*}{Advantages}\\\cline{4-7} 
&\multirow{2}{*}{}&\multirow{2}{*}{}& Seller & Buyer &Auctioneer &Commodity&\multirow{2}{*}{}&\multirow{2}{*}{} \\
\hline \hline
\cite{A.Jin2018}&Resource allocation and pricing&Efficient allocation (the number of trade maximization)&Cloudlets&MDs&Edge server &Computing resources (memory and CPU capacity)&MCC&Guarantee IR, BB and TF (IC) for both the buyers and the sellers, and CE.\\
\hline
\cite{Y.Yue2018}&Resource allocation and pricing&The system efficiency maximization (the number of trade maximization)&Edge servers&MDs&A trusted third party&Computing resources&MEC&Guarantee IR, BB and TF, improve the system efficiency, and consider the locality characteristics.\\
\hline
\cite{Y.Yue2019}&Resource allocation and pricing&The system efficiency maximization (the number of trade maximization)&Edge servers&MDs&Closest edge server&Applications&MEC&Guarantee economic properties of IR, TF and weakly BB.\\
\hline
\cite{W.Sun2018}&Resource allocation and pricing&The system efficiency maximization (the number of trade maximization)&Edge servers&IIoT MDs&Service provider&Computing resources&MEC&Guarantee BB, IR, system efficient, and TF.\\
\hline
\cite{R.Zhang2019}&Resource allocation and pricing&The system efficiency maximization (the number of trade maximization)&Edge servers and MDs&MDs&Edge server&Computing resources&MEC& Guarantee system efficiency, IR, BB and truthful properties, consider the preference of MDs and MDs can share their resources.\\
\hline
\cite{Z.Li.Z2019}&Resource-coin allocation and loan pricing&Social welfare maximization&Edge servers&IoT devices&Broker&Computing resources&Edge-assisted blockchain-enabled IoT& Guarantee IR, truthful,and BB, protect privacy information, complete fast payment, frequent trading and eliminate the impact of cold start and long return, and utilize blockchain technique.\\
\hline
\cite{Z.Li2019}&Resource allocation and pricing&Social welfare maximization&Edge servers and IoT devices&IoT devices&Broker&Computing resources&edge-cloud-assisted blockchain-enabled IoT&  Guarantee IR, truthful, IC, system efficiency, BB and protecting privacies of participants.\\
\hline
\cite{A.Zavodovski2019}&Resource allocation and pricing&Social welfare maximization (welfare loss minimization)&SPs&Clients&A distributed ledger&Computing resources&EC&Guarantee strongly BB, and IR, a secure, decentralized and truthful auctioning mechanism.\\
\hline
\cite{W.Sun2020}&Resource allocation and pricing&The system efficiency maximization&Edge servers&MDs&Trusted Edge servers&computing and storage resources&MEC& Guarantee BB, TF, and IR, and a trusted blockchain-based mechanism.\\
\hline
\cite{K.Xiao2020}&Resource allocation and pricing&The total satisfaction of users and SPs maximization&SPs&Vehicles&Blockchain-based system&Computing resources and capacities&EC& Maximize satisfaction between vehicles and SPs, improve resource utilization and a trusted blockchain-based mechanism.\\
\hline
\cite{H.Hong2020}&Task offloading&Social welfare optimization&Edge servers&MDs&MEC system&Workloads&MEC&Guarantee IR and strong BB, one task can be split into many subtasks, then assigned to different servers, and heterogeneous service prices and qualities.\\
\hline
\cite{Q.Li2020}&Resource allocation and pricing&The sum of utility maximization&MEC servers&UEs (IoT devices)&Broker&Computing resources&MEC& Combining the advantages of both RL and belief learning, outstanding convergence performance and obtain higher profit.\\
\hline
\cite{X.Peng2020}&Resource allocation and pricing&Resource utilization maximization&Fog nodes&Client vehicles&The closest BS&Computing and storage resources&VFC&Guarantee CE, IR, BB, and TF, consider nonprice attributes (location, reputation, and computing power).\\
\hline
\cite{X.Liu2019}&Resource allocation and pricing&The total utility of participants maximization&Edge server providers&MDs&Edge server&VM instances&MEC&Guarantee BB, IR and TF, higher total utility, good scalability and utilize blockchain technique.\\
\hline
\cite{S.Guo2020}&Task offloading&Enhancing the mining utility while obtaining the maximum profit&CMN and ECO&IoT MDs&Edge broker&Computing resources&Mobile blockchain network&Fully utilize available resource from idle devices in the vicinity and maximize the profit of CMN.\\
\hline
\end{tabular}
\end{table*}

\subsection{Online/Multi-Round Auction}
As an efficient and flexible form of e-commerce on the Internet, online auctions have been attracted
widespread attention from both economic and engineering fields~\cite{X.Dong2020,Y.Chen2015,B.J.Ford2013,H.S.Du2012,D.LiQ2020}.
In recently years, the applications and developments of online auctions in EC have been
accelerated by their timely responses and process
for the demands of both sellers and buyers, and the fact that the explosive growth of online
devices. Moreover, the multi-round auction also has been gained attention for allowing bidders to adopt
different strategies in the whole process. However, how to fully utilize idle
resources of MDs to assist in processing latency-sensitive and computing-intensive tasks on the edge
of the network is still worth studying, which motivates many
studies~\cite{S.Wang2017,D.Zhang2020,J.He2020,Q.Wang2019,Y.Zhang2019,C.Zhou2018}.

In~\cite{S.Wang2017}, the authors proposed an iterative auction algorithm for content placement based on a
vehicular edge computing (VEC) system. The core objective of the system is to obtain caching contents in the
least time. In the system, the contents are split into multiple content groups (CGs), which act as buyers bid
for storage resources of idle vehicles located in multiple parking lots (sellers). The VEC system plays the
role of the auctioneer to collect the bids from CGs and determine the winner. A reasonable utility function
is introduced to evaluate the values of caching the contents of each CG to each of parking lots, and then
each CG determines the bid and caching object according to the value. Based on above, the optimal placement
relation between CGs and parking lots is converted to a transmission latency minimization problem which can
be solved by the iterative ascending price auction-based caching algorithm. The algorithm can minimize the
content access latency while improving the utilities of sellers. The numerical results demonstrated validity
and effectiveness of the proposed algorithm which achieve higher the performance of average latency by around
24\% than the existing algorithms~\cite{F.Wang2012}. However, this work did not discuss the economic
properties of the proposed algorithm. Moreover, how to incentivize parked vehicles to contribute their
underutilized resources is a promising research direction toward enhancing resource utilization.

To fill the gap of~\cite{S.Wang2017}, the authors in~\cite{D.Zhang2020} utilized rewards to make resource
owners willing to share their resources. An online task offloading mechanism for EC IoT systems is designed
to achieve long-term sum-of-rewards optimization based on Lyapunov optimization and VCG auction, without
prior knowledge of the energy harvesting (EH), task arrivals or wireless channel statistics. The system
consists of IoT devices, i.e., sellers, which generate computation tasks, broadcast them and the
corresponding rewards, MDs, i.e., buyers, which bid for processing tasks to gain rewards, a task dispatcher,
i.e., an auctioneer, which collects tasks and bids from IoT devices and MDs, respectively. In addition, the
system supporting highly dynamic EH process and the randomness of tasks arrival. The EH-powered MDs evaluate
the value of task processing according to a Lyapunov optimization-based value function, and achieve long-term
sum-of-rewards optimization by utilizing Lyapunov optimization technique. A winner and payment determination
mechanism based on VCG auction is proposed to determine the winers and their payments. The optimality of
tasks assignment strategy was demonstrated by theoretical analysis and simulations. Moreover, the proposed
mechanism outperforms greedy auction and power consumption-aware auction in terms of sum-of-rewards.
However, the proposed mechanism can only guarantee TF, the other economic properties, e.g.,
IR and BB, need to be guaranteed in the future work.

Similar to~\cite{D.Zhang2020}, the authors in~\cite{J.He2020} proposed two incentive auction-based mechanisms
that both of them guarantee more economic properties. Two mechanisms include a VCG-based offline optimal
auction mechanism (VCG-OFFOAM) which owns all future information, and an online truthful auction for social
welfare maximization (ONTA-SWM) which based on the Myerson Theorem~\cite{N.Nisan2007} and can obtain the
optimal long-term social welfare without knowledge of future information. In the auction, the users and MDs
are regarded as buyers and sellers respectively, the limited computation resources of the MDs are deemed as
commodities, an MBS as an auctioneer administers the auction process. In VCG-OFFOAM, the VCG-based payment
rule can find the optimal payment and maximizes the social welfare while satisfying TF and IR. In
ONTA-SWM, the winners and corresponding payments are determined by an allocation rule based on primal-dual
technique~\cite{D.P.Williamson2007} and a payment rule. The payment rule introduces an auxiliary resource
price function to support online dynamical pricing. Both the theoretical analysis and experimental results
demonstrated that ONTA-SWM can achieve long-term social welfare maximization in polynomial time while
satisfying TF and IR. In addition, the performance of ONTA-SWM closes to VCG-OFFOAM in terms of
utility of users and the percentage of winners. However, this work did not discuss the BB of both VCG-OFFOAM
and ONTA-SWM.

Similar to~\cite{S.Wang2017,D.Zhang2020} and~\cite{J.He2020} that aim to motivate resource providers (RPs)
share their resources, the authors in~\cite{Q.Wang2019} considered the issue of incentive profit
maximization of RPs in both non-competitive and competitive scenarios based on market pricing model
and auction model. The model consists of MDs, edge clouds (RPs), computation resources and a trusted
third party, which are regarded as buyers, sellers, commodities, and an auctioneer, respectively. In
the non-competitive scenarios, they jointly consider the utility of RPs and QoE of MDs, transform
the incentive matching mechanism into a profit maximization problem based on market pricing model.
Then, a pricing scheme is designed to deal with the optimization problem where the optimal price is
found by a convex optimization method. Based on above, an online profit maximization multi-round
auction (PMMRA) mechanism for resource trading between RPs and MDs in competitive scenarios is
proposed, which achieves utility of RPs maximization while satisfying IC,
IR and efficiency. In PMMRA, the winner and the finial price are determined by
utilizing price performance ratio (PPR) and payment rule based on Vickrey
auction~\cite{W.Vickrey1961}, respectively. The experiments showed that PMMRA can obtain higher total
utility of RPs than the existing works~\cite{W.Vickrey1961,H.Zhang2017,X.Chen2019}. However, a more
practical model with mobility and randomness of MDs can be considered to be an extension.

Different from~\cite{Q.Wang2019}, the authors in~\cite{Y.Zhang2019} considered the mobility of
vehicles in designed incentive auction mechanism. They jointly consider incentives and computing
resource sharing for a smart VFC system by designing a multi-round multi-item parking reservation
auction mechanism, where the VFC system is integrated with parked vehicle assistance and smart
parking~\cite{Y.Geng2013,B.Zou2015,T.N.Pham2015,T.Lin2017}. The system consists of private parking
operators, i.e., seller, which own multiple parking places and provide parking services, moving
vehicles, i.e., buyers, which bid for parking reservation services and some of them provide
computing services, fog node constroller, i.e., an auctioneer, which manages the auction process,
and parking slots are deemed as commodities. In the system, a single-round multi-item parking
reservation auction (SMPRA) is introduced to guide the moving vehicles to desirable parking places
and motivate the parked vehicles to share computing resources. To further improve the trading price,
a multi-round multi-item parking reservation auction (MMPRA) on basic of SMPRA is proposed, which
guarantees IC, IR and BB. In MMPRA, the profit of the FNC is increased by offload pricing
update. In SMPRA and MMPRA, the allocation problem is formulated as a maximum weight perfect
bipartite matching problem and is solved by KM algorithm~\cite{H.W.Kuhn1955,J.Munkres1957}, the
payment rule improve the VCG mechanism based on Clarke pivot payments~\cite{N.Nisan2007} to
calculate payments. The simulation results demonstrated that the proposed algorithms improve the
performance of the VFC system while achieving a win-win solution for the participants.

The above related works~\cite{S.Wang2017,D.Zhang2020,J.He2020,Q.Wang2019,Y.Zhang2019} address
resource allocation problem mostly focus on enhancing resource utilization of idle resource-rich
devices. However, they do not consider the time constraint for latency-sensitive tasks.

In~\cite{C.Zhou2018}, the authors constructed a deadline-aware online resource auction (DORA)
framework to dynamically allocate computational resources more efficiently in MEC system. The system
is composed of mobile users (MUs), i.e., buyers, which need to process many computation tasks locally
or offload to the cloudlets, and an SP, i.e., a seller/auctioneer, which is equipped with limited
computational resources and hosts the auction process. In the DORA framework, the task
offloading problem is transformed as periodical auctions~\cite{X.Wang2012} that aims to achieve the
social welfare maximization. In the auction, MUs bid for computational resources (commodities) to
process a task queue, and the SP receives offloading demands from MUs, then determines the winners
of MUs, the price that the winners should pay to the SP, and allocates computational resources to
the winning MUs during each time slot. It is noteworthy that the penalty is introduced to achieve
tasks execution time constraint, i.e., complete tasks within the deadline. The DORA can obtain a
close to offline-optimal long-term social welfare with polynomial time complexity by applied
Lyapunov optimization techniques while satisfying the economic property of TF. The
simulation results showed that DORA is near-superior to some existing algorithms (e.g., HCS, LQS,
SAS)~\cite{C.Zhou2018} in terms of social welfare, average task processing delay and average task
dropping rate. The effectiveness of proposed DORA is clearly demonstrated by theoretical analysis
and the simulation results. However, the economic properties, e.g., IR or BB, need to be guaranteed
in DORA.

In summary, this section reviews the improvement of resource utilization for auction-based resource
allocation and pricing in EC. We summarize those works along with references in
Table~\uppercase\expandafter{\romannumeral4}. As shown in the table, mostly works jointly consider
the individual awareness and behavioral preferences of SPs to design online (multi-round) auction
mechanism while guaranteeing economic properties. Moreover, the proposed mechanism not only can
fully utilize idle resources but also can optimaize many performance indicators, such as social
welfare, average latency, total utility, etc. In the next section, we will discuss the use of reverse
auction approaches in EC.

\begin{table*}
\centering
\scriptsize
\caption{Online/Multi-Round Auction Mechanism in EC}
\label{table4}
\begin{tabular}{|m{0.9cm}<{\centering}|m{1.3cm}<{\centering}|m{1.5cm}<{\centering}|m{1.1cm}<{\centering}|m{1.2cm}<{\centering}|m{1.6cm}<{\centering}|m{1.5cm}<{\centering}|m{1cm}<{\centering}|m{4cm}<{\centering}|}
\hline
\multirow{2}{*}{Ref.} &\multirow{2}{*}{Issue}&\multirow{2}{*}{Objective}& \multicolumn{4}{c|}{Market structure} & \multirow{2}{*}{Scenarios} & \multirow{2}{*}{Advantages}\\\cline{4-7} 
&\multirow{2}{*}{}&\multirow{2}{*}{}& Seller & Buyer &Auctioneer &Commodity&\multirow{2}{*}{}&\multirow{2}{*}{} \\
\hline \hline
\cite{S.Wang2017}&Resource allocation and pricing&Average latency minimization&Parked vehicles&Content groups (CGs)&The VEC system&Storage resources&VEC&Guarantee well QoE, and minimize average latency.\\
\hline
\cite{D.Zhang2020}&Rewards-optimal computation offloading&Long-term sum-of-rewards optimization&IoT devices&MDs&Dispatcher&Computation resources&EC IoT&Guarantee the TF, MDs can achieve optimal utility, and support highly dynamic EH process and the randomness of tasks arrival.\\
\hline
\cite{J.He2020}&Computation offloading&Long-term social welfare optimization&MDs&Users&MBS&Computational respurces&MEC&Guarantee TF, IR and computational tractability, well QoE, the proposed algorithm can be calculated in polynomial time.\\
\hline
\cite{Q.Wang2019}&Resource allocation and pricing&Profit maximization&Edge clouds&MDs&A trusted third party &Computation resources&MEC&Advantages: Guarantee IR, IC and efficiency.\\
\hline
\cite{Y.Zhang2019}&Resource allocation and pricing&Total utility maximization&Parking place operators&On-the-move vehicles&FNC&Parking slots&VFC&Guarantee IC, IR, BB, well QoE, and achieve a win-win solution for the participants.\\
\hline
\cite{C.Zhou2018}&Resource allocation and pricing&Social welfare maximization&SPs&MUs&A trusted SP&Computational resource&MEC&The proposed algorithm can be calculated in polynomial time, guarantee TF, and achieve close-to-offline-optimal social welfare.\\
\hline
\end{tabular}
\end{table*}

\subsection{Reverse Auction}
Reverse auction~\cite{S.Parsons2011} is a popular market-based mechanisms for resource allocation which
aims to fairly allocate limited resources of SPs among multiple MDs which require a number of resources.
Recently, some works~\cite{T.H.ThiLe2020,U.Habiba2019,Q.Xu2018} apply the reverse auction-based mechanisms
to resources allocation in EC.

In~\cite{T.H.ThiLe2020}, the authors considered the problem of incentive task offloading between UEs and
vehicles in MEC, where UEs have offloading tasks requests and vehicles with computing resources
(commodities). The offloading relationship between UEs and vehicles can be well described by an auction
model, where UEs, vehicles and a BS are treated as buyers, sellers and an auctioneer, respectively. To
motivate vehicles to migrate computing tasks to the BS, a randomized auction-based incentive mechanism
is proposed which can minimize the social cost while guaranteeing TF and IR. The winners and
corresponding rewards are obtained by the randomized auction algorithm~\cite{T.H.ThiLe2020} which is
integrated with the fractional VCG auction, the decomposition algorithm~\cite{R.Carr2002,J.Li2017} and
greedy approximation algorithm. Then, the task assignment $\&$ resource allocation is transformed into
a total network delay minimization problem. They divide the problem into two subproblems: task
assignment and resource allocation, are solved by matching game~\cite{D.Gale1962,Sethuraman2006} and
convex optimization methods, respectively. The numerical results showed that the social cost of the
proposed mechanism less than renting scheme. In addition, the proposed mechanism effectively reduces
the total network delay of the system. A promising research direction is to improve the allocation efficiency.

In order to enhance computing capability of MUs, similar to~\cite{T.H.ThiLe2020}, the authors
in~\cite{U.Habiba2019} proposed a reverse auction mechanism combined with position
auction~\cite{H.Varian2007} for resource allocation and pricing in MEC offloading systems. The system
is composed of offloading users, i.e., buyers, which want to purchase computational resources
(commodities) to process computing-intensive tasks, MEC servers, i.e., sellers, which own abundant
computational resources and want to sell them, and a software-defined network (SDN) controller, i.e, an
auctioneer, which administrate the auction process. A greedy bidding strategy based on restricted
balanced bidding (RBB) algorithm~\cite{M.Cary2019} is introduced to calculate the bids of the MEC
servers which aims to maximize their utilities. The winner determination problem (WDP) is transformed into
a combinatorial optimization problem and can be tackled by an approximation algorithm in polynomial
time. In addition, the pricing mechanism utilizes the generalized second price (GSP)
auction to determine the allocation prices~\cite{H.Varian2007,K.Sowmya2013,Y.Quan2012,R.P.Leme2010}.
The theoretical analysis demonstrated that the proposed mechanism can guarantee IR, envy-free
allocation~\cite{T.Bahreini2018} and QoE of users. The numerical results exhibited the well performance
of the proposed mechanism in terms of resource utilization of the system and QoE satisfaction of MUs.
However, the fewer MUs may lead to lower efficiency of the system.

The approaches~\cite{T.H.ThiLe2020} and~\cite{U.Habiba2019} do not consider the reliability and
security of SPs, which may cause low allocation efficiency and even privacy leakage. Therefore, the
authors in~\cite{Q.Xu2018} designed a trustworthy caching and bandwidth allocation scheme based on
reverse auction for MUs in the mobile social networks (MSNs). A trust evaluation model includes two
part, i.e., direct evaluation and indirect evaluation, is constructed to help an MUs (buyer) to
identify the reliability of nearby edge nodes (sellers) and find a better one to cache the contents. The
direct evaluation is according to historical interactions and indirect evaluation is according to the
suggestions from other MUs. Then, the optimal edge node for each MUs can be determined by the reverse
auction algorithm. The Bayesian equilibrium is found by back induction method to determine the optimal
caching space and bandwidth allocation for MUs. The proposed scheme can guarantee well QoE of MUs while
preventing malicious edge nodes from attacking MUs. However, a more secure and reliable model that
prevents participants from privacy leakage~\cite{Z.Li2019,A.Zavodovski2019,W.Sun2020,K.Xiao2020} needs
to be investigated in the future work.

In this section, we review the existing works of reverse auction-based resource allocation and pricing
approaches in EC. As shown in Table~\uppercase\expandafter{\romannumeral5}, we can easily find that the
reliability of SPs and QoE/QoS of MUs have been attracting more and more attention in designing resource
allocation mechanisms. However, many economic properties, e.g., TF and IR, need to be considered in the
future work. In the next section, we will discuss the use of hierarchical auction approaches in EC.

\begin{table*}
\centering
\scriptsize
\caption{Reverse Auction Mechanism in EC}
\label{table5}
\begin{tabular}{|m{0.9cm}<{\centering}|m{1.3cm}<{\centering}|m{1.5cm}<{\centering}|m{1.1cm}<{\centering}|m{1.2cm}<{\centering}|m{1.6cm}<{\centering}|m{1.5cm}<{\centering}|m{1cm}<{\centering}|m{4cm}<{\centering}|}
\hline
\multirow{2}{*}{Ref.} &\multirow{2}{*}{Issue}&\multirow{2}{*}{Objective}& \multicolumn{4}{c|}{Market structure} & \multirow{2}{*}{Scenarios} & \multirow{2}{*}{Advantages}\\\cline{4-7} 
&\multirow{2}{*}{}&\multirow{2}{*}{}& Seller & Buyer &Auctioneer &Commodity&\multirow{2}{*}{}&\multirow{2}{*}{} \\
\hline \hline
\cite{T.H.ThiLe2020}&Task assignment and resource allocation&Alleviate high load of BS and minimize total delay of all UEs&Vehicles&UEs&BS&Computation resources&MEC&Guaranteeing TF and IR, minimize the social cost and network delay.\\
\hline
\cite{U.Habiba2019}&Resource allocation and pricing&The utility maximization&CSPs&Offloading users&SDN controller&Computational resources&MEC&Guarantee envy-free and IR, well Qos, and ensures users' satisfaction.\\
\hline
\cite{Q.Xu2018}&Resource allocation and pricing&The utility maximization&Edge nodes&MUs&The mobile user&Caching space and bandwidth resources&EC&Protect the network from the attacks of malicious edge nodes, and ensure well QoE.\\
\hline
\end{tabular}
\end{table*}

\subsection{Hierarchical Auction}
More recently, hierarchical strategies have been widely applied to study the distributed systems, e.g.,
MBNs~\cite{Y.Jiao2018}, AI~\cite{E.Sirin2004}, 5G~\cite{M.Chen2020} and robotics~\cite{J.Oksanen2015}.
The hierarchical auctions as an emerging approach can be applied to deal with the issues
of resource allocation while achieving the social welfare maximization of the system.

To motivate the edge servers to provide services for MUs while ensuring security of the
trading system, the authors in~\cite{C.Xia2018} proposed an incentive efficient three-stage auction
mechanism (ETRA) for resource allocation in an MBN. The MUs (buyers) as miners have the same wireless network
access point (auctioneer) defined as a group, and then compete with each other to offload mining tasks
to edge servers (sellers) with computation resources (commodities). A group-buying scheme is designed
to help miners to afford resources and incentivize edge servers to participate in resource trading. The
ETRA consists of three stages, i.e., matching potential winner, matching cloudlet for AP and allocation
the resource. In the first stage, the miners submit bids and resource requests to APs according to
their preference for distance and QoS, then the potential winners and their payments can be calculated
by the payment calculation algorithm based on VCG mechanism. In the second stage, APs find out the
optimal matching between MUs and edge servers. In the last stage, the requested resources are placed in
corresponding APs and are allocated to the miners. The theoretical analysis demonstrated that ETRA
guarantee TF, IR and CE while achieving the social
welfare maximization. The simulation results showed that the performance of the ETRA outperforms
TACD~\cite{G.Zhou2017} and HAF~\cite{M.Jia2017} in terms of utility of miners and social welfare.
However, the computational complexity was not discuss in this work.

In this section, we discusses the problem of resource allocation by an incentive hierarchical auction-based
mechanism in EC. The mechanism jointly considers the network economics and individual preferences of SPs
to efficiently allocate resources for MUs. In the next section, we will discuss the use of revenue-optimal
auction approaches in EC.

\subsection{Revenue-Optimal Auction}
It is still a challenge to fully utilize the resources of SPs to process computing-intensive tasks of
MDs due to SPs have individual awareness and behavioral preferences. Recently, some
works~\cite{D.Bermbach2020,Q.WangS2019,A.Kiani2017,K.Zhu2020,N.C.Luong2018,N.C.Luong2020} are motivated
to address the problem via maximizing the revenue of SPs while aiming to satisfy the desirable economic
properties, e.g., IC and IR.

In~\cite{D.Bermbach2020}, the authors addressed the problem of function allocation based on a distributed
auction-based mechanism in fog-based FaaS platforms. In the platform, the application developers act as
buyers to bid for computing and storage resources (commodities), and the service nodes, which play the
different roles include sellers and auctioneers contain three types: edge, intermediary and cloud, which
offer computing and storage services for applications. In the auction, developers submit two types of bids,
i.e., storage bids and processing bids, then service nodes reject or accept the requests according to
their remaining resources. Particularly, the cloud nodes will accept all requests which are rejected by
non-cloud nodes. In addition, they utilize the first-price auction to determine the final price. The
simulation results demonstrated that the proposed mechanism can achieve revenue of service nodes maximization
while stisfying all demands of application developers. Although the auction-based mechanism is simplistic, it
opens a valuable research direction toward function placement in fog-based FaaS platforms. Moreover, a
more practical model that jointly considers location of service nodes and available resources of
clients can be investigated in the future work.

In contrast to~\cite{D.Bermbach2020}, the authors in~\cite{Q.WangS2019} placed more attention on
incentives for resource trading in MEC, they achieved profit maximization of SPs by utilizing
market-based pricing model. Similar as~\cite{Q.Wang2019}, this work also considered incentive algorithms
of SPs both in non-competitive and competitive scenarios. In the non-competitive scenarios, the utility
of SPs and the constraint on each user gain are jointly considered to establish a matching relationship
between edge servers (sellers) and MDs (buyers) based on market-based pricing model. Then, apply it to design
a profit maximization multi-round auction (PMMRA) algorithm in competitive scenarios which guarantees IC,
IR and CE. The PMMRA algorithm can be calculated in polynomial time and consists of three crucial roles,
i.e., bidding strategy, winner determination and payment determination. The first role evaluate which
seller is better for buyers based on calculated the bid performance ratio (BPR). The second role determines
the SP with higher price performance ratio (PPR) can be served by the SP. The final price can be calculated
by Vickrey auction~\cite{W.Vickrey1961} that adopted by the last role. The simulation results showed that the performance of
PMMRA outperforms the algorithms of~\cite{H.Zhang2017} and~\cite{X.Chen2019} in terms of the utility of
SPs and the number of offloaded tasks, respectively. However, a more effective incentive scheme which
considers MDs have individual awareness and behavioral preferences can be considered to be an
extension.

Different from~\cite{D.Bermbach2020,Q.WangS2019}, the authors in~\cite{A.Kiani2017} introduced a hierarchical
MEC (HI-MEC) architecture which is inspired by the principles of LTE-advanced backhaul network and includes
three levels, i.e., field, shallow, and deep cloudlets. To achieve profit maximization of the service
provider, the supply-demand relation between edge-computing service provider (SP) and MUs can be formulated
as an auction where MUs act as buyers, the VMs are deemed as commodities with diverse types which contain
computing and communications resources, and the SP plays different roles include a seller and an auctioneer.
Then, a two time scale optimization algorithms for resource allocation are proposed based on HI-MEC.
Specifically, the auction-based profit maximization of the SP for VM pricing and VM assignment is transformed
into a binary linear programming (BLP)~\cite{D.Justice2006} problem and then solved by the heuristic
algorithms. Note that the prcie depends on the number of demands and the available resources. To achieve the
total network delay minimization, bandwidth allocation problem is formulated as a convex optimization problem
and then addressed by a centralized optimal solution. The simulation results demonstrated validity and
effectiveness of the proposed algorithms based on HI-MEC architecture. However, this work only supported a user
request a single type VM at a time. A more efficient auction-based model~\cite{T.Bahreini2018} that allow a
user requests a bundle of VMs of diverse types needs to be investigated in the future work.

\begin{table*}
\centering
\scriptsize
\caption{Revenue-Optimal Auction Mechanism in EC}
\label{table6}
\begin{tabular}{|m{0.9cm}<{\centering}|m{1.3cm}<{\centering}|m{1.5cm}<{\centering}|m{1.1cm}<{\centering}|m{1.2cm}<{\centering}|m{1.6cm}<{\centering}|m{1.5cm}<{\centering}|m{1cm}<{\centering}|m{4cm}<{\centering}|}
\hline
\multirow{2}{*}{Ref.} &\multirow{2}{*}{Issue}&\multirow{2}{*}{Objective}& \multicolumn{4}{c|}{Market structure} & \multirow{2}{*}{Scenarios} & \multirow{2}{*}{Advantages}\\\cline{4-7} 
&\multirow{2}{*}{}&\multirow{2}{*}{}& Seller & Buyer &Auctioneer &Commodity&\multirow{2}{*}{}&\multirow{2}{*}{} \\
\hline \hline
\cite{D.Bermbach2020}&Resource allocation and pricing&Revenue maximization&Service nodes&Application developers&Service nodes&Computing and storage resources&FC&Maximize revenue of service nodes, and opens a valuable research direction toward function placement in fog-based FaaS platforms, and incentive mechanism.\\
\hline
\cite{Q.WangS2019}&Task offloading&Profit maximization&Edge servers&MDs&A trusted third party&Computational resources&MEC&PMMRA has polynomial-time complexity, and consider incentive algorithms of SPs both in non-competitive and competitive scenarios.\\
\hline
\cite{A.Kiani2017}&Resource allocation and pricing&Profit maximization&ECSPs&MUs&ECSPs&VM instances (Computing and communications resources)&MEC&Facilitate the resource allocation for MEC networks, and well QoE.\\
\hline
\cite{N.C.Luong2018}&Resource allocation and pricing&Profit maximization&SP&MUs&SP&Edge computing resource units&EC&Guarantee IC, IR, combine three hot technologies: EC, DL and blockchain, and well QoS.\\
\hline
\cite{N.C.Luong2020}&Resource management and pricing&Profit maximization&Fog nodes&User devices&SP&Fog computing resource units&FC&Guarantee IC, IR, TF, combine three hot technologies: FC, DL and blockchain, and well QoS.\\
\hline
\end{tabular}
\end{table*}

In recent years, machine learning (ML) as a promising technology opens a new research direction toward
addressing optimization problems on resource allocation of auction market~\cite{N.C.Luong2018,N.C.Luong2020}.

To support blockchain applications in mobile environments, the authors in~\cite{N.C.Luong2018} utilized deep
learning (DL) techniques to design the optimal auction for edge resources allocation to achieve
blockchain-assisted EC. The model is composed of one SP, i.e., the seller and the auctioneer, which owns
limited computing resources (commodities), multiple MUs, i.e., buyers, which play the role of miners and
compete with each other for one computing resource unit by submitting bids. It is novel to create the neural
networks to implement the allocation and payment rules based on an analytical solution of the optimal
auction~\cite{R.B.Myerson1981}, and then calculating the winning probability of miners and their payments
according to the inputs of neural networks, i.e., miners' bids. Therefore, the auction mechanism optimized by
the neural networks, which is trained with valuations of the miners, to maximize the revenue of the SP while
ensuring IC and IR. The experimental results verified that the proposed mechanism can obtain higher profit of
the SP than the traditional sealed-bid auction~\cite{W.Vickrey1961}. However, this work only supported a
single resource unit and did not consider the privacy protection of bidders.

As an extension of~\cite{N.C.Luong2018}, the auction-based mechanism in~\cite{N.C.Luong2020} can support
buyers to buy more than one resource unit in each auction, i.e., an optimal auction mechanism for resource
allocation via leveraging deep learning in FC, which can achieve revenue maximization of SPs and ensure well
QoS of users. The market consists of fog nodes, i.e., sellers, which owns a certain amount of computing
resource units, miners, i.e., the buyers, that are lightweight devices bidding for computing resources to
solve Proof-of-Work (PoW) puzzles, and one service provider, i.e., the auctioneer, that determines the
resource assignments and the payments. The winners and payments are decided by the assignment and payment
systems respectively, both of which are deep learning systems. The outputs of assignment and payment systems
are assignment probabilities of miners and corresponding prices that winning miners need to pay,
respectively. The simulation results demonstrated that the proposed mechanism is superior to the greedy
algorithm~\cite{Y.Akcay2007} in terms of revenue, IC and IR violations. In addition, the TF
is guaranteed by the mechanism. However, a worthy work of maximizing social welfare while satisfying IC, IR
and TF needs to be investigated in the future work.

In summary, this section reviews the existing works which improve the allocation efficiency from the
perspective of revenue optimization. These works along with their references are summarized in
Table~\uppercase\expandafter{\romannumeral6}. As shown in the table, the profit-driven nature of participants
and well QoS of users are considered in designing incentive auction-based mechanisms to stimulate SPs to
offer services for MDs. We observe that the decentralized and hierarchical auction architectures are
considered for resource allocation in the works~\cite{D.Bermbach2020,A.Kiani2017}. In addition, some works
study the combination of auction theory with DL to address resource allocation in blockchain
networks~\cite{N.C.Luong2018,N.C.Luong2020}. Generally, how to achieve revenue maximization while satisfying
diverse requirements still needs to be investigated in the future work.




\section{Open Issues and Future Research Directions}
As we have discussed in aforementioned sections, auction theory provides an effective solution to address
various issues for EC, e.g., resource allocation, pricing scheme, task offloading, energy consumption optimization etc. In addition to the existing works, there are still several open challenges and promising
research directions for EC are summarized as follows.
\subsubsection{Security and Privacy}
As discussed in~\cite{F.Zhang2018,Y.Jiao2018,Y.Jiao2019,Y.Yue2018,X.Liu2019,Q.Wang2019,C.Zhou2018},
auction-based mechanisms commonly rely on the third parties, (e.g., edge servers) to administrate each
auction process. However, it is inevitable to cause the issues of security and privacy leakage~\cite{W.Sun2020,C.Xia2018}. In order to deal with this problem, many existing works~\cite{W.Sun2020,K.Xiao2020} utilize blockchain technique to
protect trading information from tampering by malicious third parties. However, an auction algorithm
is executed in a blockchain network needs to share their private information to other users within the same
blockchain network which cannot avoid potential privacy leakage~\cite{K.Xiao2020}. Therefore, the open issue
is how to ensure security and reliability of the auction-based mechanisms while preventing from privacy leakage~\cite{A.Zavodovski2019,U.Habiba2018}. In addition, it is crucial to prevent malicious edge servers from attacking MDs~\cite{Q.Xu2018}.
\subsubsection{Incentive Scheme}
As a promising computing architecture, EC addresses the problem of the insufficient ability in
processing computing intensive and latency-sensitive tasks of the MDs by utilizing idle resources of edge servers or UEs. However, SPs (edge servers or UEs) and SRs (MDs or users) are not always consistent in their interests due to
their selfishness~\cite{S.Pan2019}. Many works~\cite{W.Sun2020,T.H.ThiLe2020,Q.WangS2019,Y.Liu2017,Z.Gao2019}
propose incentive auction-based mechanisms for EC in resource allocation~\cite{W.Sun2020}, task
offloading~\cite{T.H.ThiLe2020} and etc. However, the proposed mechanisms cannot fully utilize idle-resources
of SPs due to individual awareness and behavioral preferences of both SPs and SRs. Thus, it is crucial to
design incentive auction-based mechanisms according to the preferences and the conflicting interests of both
sides of trading to stimulate them to willing participate in the resource trading.
\subsubsection{Solutions of WDP}
Almost all existing works solve the winner determination problem (WDP) generally by applying mathematical
optimization techniques~\cite{Y.Hung2018,Y.Hung2020}. However, it is hard to attain the optimal solutions of
the WDP because it belongs to NP-complete problems and is inapproximable~\cite{T.W.Sandholm1996}. Recently, machine learning (ML) plays an increasingly crucial role in addressing various NP-hard
problems~\cite{M.Lee2018,M.Prates2019,M.Lee2020}. Thus, auction theory combined with ML can be a prominent
research direction to solve the WDP.
\subsubsection{Combination of Auction Theory with Federated Learning}
In order to make IoT devices more intelligent, federated learning (FL)~\cite{Q.Yang2019} is applied to train
a shared model by multiple edge nodes cooperating with each other, and then deploy the model to required IoT
devices. Therefore, FL is a suitable and promising ML technology to shorten model training time while
preserving data privacy of users in EC. Recently, many works place main attention on enhancing the
performance of FL algorithms~\cite{X.WangC2020} privacy and security problems~\cite{Y.Lu2020}. However, these
work mostly assume that edge nodes are willing to participate in computing without any returns. In real
world systems, edge nodes cannot offer resources unconditionally due to their limited resources, and
have different preferences for different computing requests. Auction as an efficient method
can be leveraged to design incentive mechanisms to motivate edge nodes contribute their
resources~\cite{A.Bandyopadhyay2020,A.Jin2018,A.Zavodovski2019,Q.Wang2019,Y.Zhang2019,T.H.ThiLe2020}.
This opens a promising research direction toward integrating auction approaches towards FL process~\cite{S.Fan2020}.

In conclusion, the combination of auction theory with blockchain and ML applying to EC will become an obvious research trend.
\section{Conclusions}
In this survey, we have comprehensively introduced and discussed recent work of the applications of
auction-based mechanisms for EC. Firstly, we have introduced the main paradigms of EC (i.e., cloudlets,
FC, and MEC) and key advantages of each computing paradigm. Then, we have presented the related terminologies
of auction theory and given a brief introduction of related auction methods. After that, we have presented
detailed reviews, analyses and comparison of the approaches exploiting auction-based models to solve
various resource allocation related issues in EC. Finally, several open challenges and promising
research directions have indicated. In conclusion, we hope this paper can provide clear guidance for
researchers who is interesting in applying auction approaches for EC.

\bibliographystyle{IEEEtran}
\bibliography{cite}

\end{document}